\documentclass[prx,reprint,superscriptaddress]{revtex4-1}
\usepackage{amsmath,amssymb,bm,mathrsfs,graphicx, braket,amsthm,times}
\usepackage[colorlinks=true,citecolor=blue,linkcolor=blue]{hyperref}
\usepackage[normalem]{ulem}
\usepackage[usenames,dvipsnames]{color}
\usepackage{multirow} 
\usepackage{soul}

\begin{document}

\title{Fragile topological phases in interacting systems} 
\author{Dominic V. Else}  
\affiliation{Department of Physics, Massachusetts Institute of Technology, Cambridge, MA 02139, USA}
\affiliation{Department of Physics, University of California, Santa Barbara, CA 93117, USA}

\author{Hoi Chun Po}  
\affiliation{Department of Physics, Massachusetts Institute of Technology, Cambridge, MA 02139, USA}
\affiliation{Department of Physics, Harvard University,
Cambridge, MA 02138, USA}

\author{Haruki Watanabe}  
\email{haruki.watanabe@ap.t.u-tokyo.ac.jp}
\affiliation{Department of Applied Physics, University of Tokyo, Tokyo 113-8656, Japan.}

\newcommand{\U}{\mathrm{U}}

\begin{abstract}
Topological phases of matter are defined by their nontrivial patterns of ground-state quantum entanglement, which is irremovable so long as the excitation gap and the protecting symmetries, if any, are maintained.
Recent studies on noninteracting electrons in crystals have unveiled a peculiar variety of topological phases, which harbors nontrivial entanglement that can be dissolved simply by the the addition of entanglement-free, but charged, degrees of freedom. Such topological phases have a weaker sense of robustness than their conventional counterparts, and are therefore dubbed ``fragile topological phases.'' 
In this work, we show that fragile topology is a general concept prevailing beyond  systems of noninteracting electrons. Fragile topological phases can generally occur when a system has a $\mathrm{U}(1)$ charge conservation symmetry, such that only particles with one sign of the charge are physically allowed (e.g.\ electrons but not positrons).
We demonstrate that fragile topological phases exist in interacting systems of both fermions and of bosons.
\end{abstract}

\maketitle

\begin{table*}[t]
\center
\caption{Summary of phases of intermediate stability}
\begin{tabular}{c|ccccc}
\hline \hline
Class & $\,\,\,$Trivial$\,\,\,$ & $\,\,\,$Obstructed trivial$\,\,\,$ & $\,\,\,$Fragile topological$\,\,\,$  & $\,\,\,$Stably topological$\,\,\,$  \\\hline
Nontriviality&Trivial &   \multicolumn{2}{c}{\begin{minipage}{70mm}\centering\scalebox{0.50}{\includegraphics{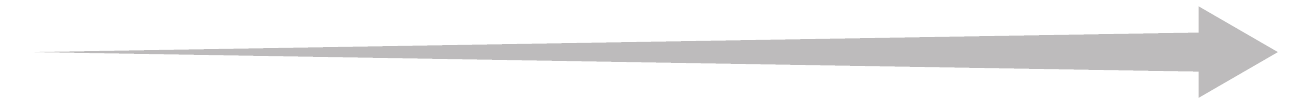}}\end{minipage} }  & Topological \\
Trivialized by & --   & uncharged ancillas & charged ancillas \\
(interpretation) & -- & (adding new ions or orbitals)  & (adding valence electrons) & --  \\
Examples & Product States & Secs.~\ref{sec:Square}--\ref{sec:bosons} &  Secs.~\ref{sec:Square}--\ref{sec:bosons} &Chern, $\mathbb{Z}_2$ TI\\
\hline \hline
\end{tabular}
\label{tab:summary}
\end{table*}

\section{introduction}
A fundamental problem in condensed matter physics is to identify and classify non-trivial topological phases of matter~\cite{Zak2000, Schnyder, Kitaev, 
Pollmann2010,Ryu2010,Pollmann2012,Fidkowski2010,Schuch_1010,Chen2010,
Fidkowski2011,CGLW,Chen2011b,MooreFreed,Slager2012,
Vishwanath2013,Molenkamp, Wang2014, K,Gu2014,Burnell2014,
Wang2015,Oshikawa_SPt,Cheng2015,Gomi, Song_1604,Combinatorics, Thorngren_1612,Ken2017,NC, Bradlyn17,Huang_1705},
 including in the presence of symmetries. But what, precisely, do we mean by ``non-trivial''? In this paper, we will revisit this question, and discuss phases of matter whose characterization as ``trivial'' or ``non-trivial'' depends rather sensitively on the precise definitions used.

Traditionally, we say that a system with a gapped local Hamiltonian $\hat{H}$ is in a non-trivial topological phase if there is no smooth deformation of local Hamiltonians, preserving the gap and all relevant symmetries, that relates $\hat{H}$ to a different Hamiltonian $\hat{H}'$ for which the ground state is simply a product state. But we still have to specify the space of Hamiltonians which this deformation is allowed to pass through. Specifically, we have to specify whether the Hamiltonians must act only on the degrees of freedom of the original Hamiltonian $\hat{H}$, or whether, along the path, we are allowed to introduce additional degrees of freedom, initially in a product state with each other and with the original degrees of freedom.
We will refer to such additional degrees of freedom as ``ancillas.'' In the context of tight-binding models of fermions, introducing ancillas is equivalent to introducing additional ions and/or additional orbitals into the tight-binding description.

Can the non-triviality of a phase depend on whether or not ancillas are permitted? For phases with purely internal symmetries, there is an argument that it cannot, because even if ancillas are not explicitly permitted, one can build effective ancillas from the unused degrees of freedom after coarse-graining~\cite{Schuch_1010}. This argument fails, however, for systems with crystalline symmetries, which cannot be coarse-grained while preserving the symmetries. This raises the possibility of phases of ``intermediate'' stability that are non-trivial in the absence of ancillas, but become trivial in their presence. 

In this paper, we will focus on systems composed of particles carrying a conserved $\U(1)$ charge, such that the charges of all the particles are all of the same sign. A canonical example of this setting is an electronic problem with charge conservation. We will also always assume lattice translation symmetries, such that the filling $\nu$, defined as the average charge per unit cell, is a well-defined conserved quantity.

We will discuss two kinds of phases of intermediate stability in this setting.
The first we refer to as a \emph{fragile topological phase}~\cite{fragile, fragileBAB1, Robert, fragileBAB2}. A fragile topological phase remains non-trivial even in the presence of ancillas as long as the filling $\nu$ is kept unchanged.
However, if we introduce {\it charged} ancillas which increase the total filling, then the resulting state can be trivialized. In the language of electronic tight-binding models, the state is trivialized if we add \emph{occupied} orbitals, i.e. we add valence electrons, where the electrons added are initially in a trivial insulating state.
The second, which is more trivial by comparison, we refer to as an \emph{obstructed trivial phase}~\cite{Bradlyn17,PhysRevB.97.035139}. Similar to topological phases, obstructed trivial phases showcase symmetry-protected quantum entanglement. However, such entanglement is protected only when one stays strictly with the original degrees of freedom, and so state can be trivialized as soon as we incorporate {\it uncharged} ancillas. In the language of tight-binding models, these uncharged ancillas can come from {\it unoccupied} orbitals, which are always present when one recalls the fact that the electrons live in the continuum.

Fragile topological phases~\cite{fragile, fragileBAB1, Robert, fragileBAB2} and obstructed trivial phases~\cite{Bradlyn17,PhysRevB.97.035139} have previously been discussed in the context of free-fermion systems. In this paper, we will show that these notions are more general and also apply in the presence of strong interactions. For example, we will show that some fragile topological phases identified in free-electron systems are robust to interactions. An important step in our argument is the introduction of charge carriers with the opposite sign, which we dub ``positrons'' by analogy. Introducing positrons violates the ``single-sign'' assumption we imposed on the charge carriers, and we will show that a fragile topological phase can generically be trivialized upon the lifting of this assumption. Moreover, we will sketch a construction of an \emph{intrinsically interacting} fragile topological phase in a bosonic system.  Our arguments are based on recent results on the classification of topological phases with spatial symmetries \cite{Song_1604,Thorngren_1612,Huang_1705,ElsePreparation}, which in the cases discussed here reduces to ``lattice homotopy'' \cite{Zak2000,LattHomotopy,Bradlyn17}.

This paper is organized as follows. 
In Sec.~\ref{sec:definitions}, we present the precise definition of fragile topological phases and obstructed trivial phases.
We then present in Sec.~\ref{sec:Square} a general argument on the stability of certain fermionic fragile topological phases and obstructed trivial phases against the introduction of interactions, through a detailed example of spinless electrons defined on an inversion-symmetric square lattice.
The discussion is extended to spinful electrons (Sec.~\ref{sec:honeycomb}) and hard-core bosons (Sec.~\ref{sec:bosons}) on the honeycomb lattice. We discuss the particle-hole duality between the obstructed trivial phase and the fragile topological phase in Sec.~\ref{sec:duality} and then conclude in Sec.~\ref{sec:conclusion}.

\section{Definitions of phases of intermediate stability}
\label{sec:definitions}
As we have alluded to, in the presence of spatial symmetries the apparent dichotomy between trivial and topological phases is more subtle than it may appear, due to the existence of ``phases of intermediate stability'' whose stability is dependent on the admittance of ancillas. To understand the properties of fragile topological phases, it will be beneficial to first provide precise definitions for the various phases involved, which we summarize in Table~\ref{tab:summary}. We will always assume that the full symmetry group $G$ includes U(1) symmetry and the $d$-dimensional lattice translation symmetry.

\subsection{Setups and Rules}
\label{sec_ancillas}
As a preparation, let us first clarify the setup.  
\subsubsection{Local Hilbert space}
\label{sec:hilbert}
Suppose that the model of our interest is defined on a lattice $\Lambda\subset \mathbb{R}^d$ symmetric under $G$. 
For each $x \in \Lambda$, let $G_{x}$ be the subgroup of $G$ that leaves $x$ unmoved. 
States in the local Hilbert space $\mathcal{H}_{x}$ can be classified accordingly by the irreducible representations of $G_{x}$.  In particular, $G_x$ includes the $\U(1)$ symmetry $e^{i\theta \hat{Q}_x}$ that defines the local U(1) charge. The entire Hilbert space $\mathcal{H}_\Lambda$ is given by the tensor product $\otimes_{x\in\Lambda}\mathcal{H}_{x}$. 

We demand that the charge operator $\hat{Q}_x$ satisfies the positive-semidefinite condition $\hat{Q}_x \geq 0$ on $\mathcal{H}_x$. Namely, any eigenstate of $\hat{Q}_x$ in the local Hilbert space has a nonnegative eigenvalue of $\hat{Q}_x$.  In electronic systems, this condition is violated when positronic states are allowed. (Here, and throughout this paper, we define electrons to have positive charge and positrons to have negative charge.)

We also assume that the local Fock vacuum $|0\rangle_{x}$ is the only charge-0 state in the local Hilbert space $\mathcal{H}_{x}$ and that $|0\rangle_{x}$ has the trivial representation $U_{x}(g)=1$ for all elements $g\in G_{x}$.

\subsubsection{Ancillas}
\label{sec:ancillas}
In our definition of equivalence of topological phases, we will allow ourselves to add ancillas.  To define this operation properly, let $y_0\in\mathbb{R}^d$ be a point that may not belong to $\Lambda$.   The orbit $\{g(y_0)\,|\,g\in G\}$ defines a $G$-symmetric lattice $\Lambda'\subset \mathbb{R}^d$.  We introduce a new local Hilbert space $\mathcal{H}_y$ to each $y\in\Lambda'$.  We assume the same conditions on $\mathcal{H}_y$ as mentioned for $\mathcal{H}_x$ in the previous section.

Let us choose $|\phi\rangle_{y_0}\in\mathcal{H}_{y_0}$ that obeys a one-dimensional representation of $G_{y_0}$ (as is necessary if we want to avoid introducing ground state degeneracy when introducing the ancilla). We assume that the symmetry image $\hat{g}(|\phi\rangle_{y_0})$ belongs to $\mathcal{H}_{g(y_0)}$.  Then one can construct a $G$-symmetric product state $|\phi\rangle_{\Lambda'}$ by
\begin{equation}
\label{eq:ancillas}
|\phi\rangle_{\Lambda'}=\otimes_{g\in G/G_{y_0}}\hat{g}(|\phi\rangle_{y_0}),
\end{equation}
which belongs to $\mathcal{H}_{\Lambda'}\equiv\otimes_{y\in\Lambda'}\mathcal{H}_{y}$.

What we mean by introducing ancillas is the following. Let $|\Psi\rangle_{\Lambda}\in\mathcal{H}_\Lambda$ be the ground state with a finite excitation gap. Then after introducing ancillas, the total Hilbert space is enlarged to $\mathcal{H}_\Lambda\otimes \mathcal{H}_{\Lambda'}$ and the ground state becomes $|\Psi\rangle_{\Lambda}\otimes|\phi\rangle_{\Lambda'}\in\mathcal{H}_\Lambda\otimes \mathcal{H}_{\Lambda'}$.

In this paper, we will consider different kinds of ancillas, which will lead to different notions of phases. Firstly, we can distinguish between \emph{electronic} ancillas and \emph{general} ancillas. An electronic ancilla is one in which the Hilbert space $\mathcal{H}_{y_0}$ satisfies the same conditions (i.e.\ all states are charge posisitive-semidefinite, and the vacuum state is unique and carries trivial representation of the on-site symmetry) as we imposed on the Hilbert space $\mathcal{H}_x$ of the original lattice degrees of freedom in the previous section. A non-electronic ancilla, meanwhile, violates these conditions. Non-electronic ancillas seemingly have little physical relevance for condensed matter systems, but we will employ them as a theoretical device.

Secondly, we distinguish between \emph{charged} and \emph{uncharged} ancillas, depending on the properties of the state $\ket{\phi}_{y_0}$ in which the ancilla is in when it is added. We say the ancilla is uncharged if $\ket{\phi}_{y_0}$ carries trivial representation of the on-site symmetry group, $U_{y_0}(g) \ket{\phi}_{y_0} = 1$. In particular, this implies that the $\mathrm{U}(1)$ charge must be zero. For electronic ancillas, the latter is also a \emph{sufficient} condition for the ancilla to be uncharged,  since it implies that $\ket{\phi}_{y_0}$ must be the Fock vacuum $\ket{0}_{y_0}$.

Now to define fragile topological phases, we allow only ancillas that are \emph{electronic} and \emph{uncharged}; in particular, such ancillas do not change the filling $\nu$. Although such ancillas are always initially in the Fock vacuum state, they can still help in trivializing the ground state, since once the ancillas are introduced the ground state is allowed to explore the enlarged Hilbert space $\mathcal{H}_\Lambda\otimes \mathcal{H}_{\Lambda'}$ and states other than $|0\rangle_{y_0}$ in $\mathcal{H}_{y_0}$ may be utilized upon smooth deformation.

One might ask what is the physical reason for considering such ancillas. Let us focus on the case of tight-binding models of electrons. The point is that tight-binding models are based on choosing a finite number of orbitals centered at each lattice site (thus giving a finite-dimensional site Hilbert space). Thus, one can think of the formal device of an ``ancilla'' simply as a way to bring in additional orbitals that were originally left out of the tight-binding approximation. This makes it clear why we want the ancillas to be uncharged: the process of bringing in an ancilla, though it seems abrupt in terms of the finite-dimensional site Hilbert space, is really a change in our \emph{description} of the system rather than the underlying system itself; this could never be true for a process that introduces extra charges.

However, there is still a physical interpretation for the process of adding a \emph{charged} electronic ancilla. Usually in solid state physics, we divide electrons into core electrons which are tightly bound to their respective nuclei, and the valence electrons which determine the low-energy properties of the solid. However, in the course of a deformation it is possible that that wavefunctions of the core electrons could become less localized, to the point where we have to start treating them as valence electrons. These core electrons, and the orbitals they occupy, then effectively ``enter the scene'' for our description of the material. This corresponds to adding a charged ancilla. To the extent that we classify phases with the restriction of uncharged ancillas, we are restricting ourself to the regime where the core electrons, if they exist, always remain tightly bound.

\subsection{Definitions of phases}
Now we are ready to state the definitions of phases in Table~\ref{tab:summary} one by one.
\subsubsection{Trivial phases}
The ground state $|\Psi\rangle_{\Lambda}$ is trivial when it is smoothly deformable to a product state in $\mathcal{H}_{\Lambda}$, i.e., without introducing any ancillas.

\subsubsection{Obstructed trivial phases}
The gapped ground state $|\Psi\rangle_{\Lambda}$ is in an obstructed trivial phase if it is not trivial without ancillas, but becomes trivial when uncharged electronic ancillas are introduced. Examples are discussed in Secs.~\ref{sec:Square}--\ref{sec:bosons}. As we will later see, their existence is closely related to the observation that point-group symmetries can lead to mutual distinction between trivial phases~\cite{Zak2000, Oshikawa_SPt, Combinatorics, NC, Bradlyn17}. 

\subsubsection{Fragile topological phases}
The ground state $|\Psi\rangle_{\Lambda}$ is in a fragile topological phase if (i) it cannot be trivialized by introducing any uncharged electronic ancillas, but (ii) can be trivialized by adding charged electronic ancillas. This definition generalizes  noninteracting ``fragile topological insulators" (FTIs)~\cite{fragile, fragileBAB1, Robert, fragileBAB2}. 
Examples are discussed in Secs.~\ref{sec:Square}--\ref{sec:bosons}.

\subsubsection{Stably topological phases}
The ground state $|\Psi\rangle_{\Lambda}$ is in a stably topological phase if it cannot be trivialized by adding any electronic ancillas, even if they are charged. (Note that one can easily convince oneself that allowing also non-electronic ancillas would not make any difference to this definition.)

\subsubsection{Comparison with previous definitions}
It is useful to compare the above definitions with previous works on the classification of interacting topological phases with spatial symmetries \cite{Song_1604,Thorngren_1612,Huang_1705,ElsePreparation}. In such works, phases are classified with respect to deformations that can include ancillas that need not be electronic, but must be uncharged. (In general, there can be more than one ``trivial'' phase under such an equivalence relation). Such an equivalence relation is not very physically relevant, as we have argued. Nevertheless, these results will still form the starting point for our analysis of the more physically motivated definitions given above.

Lastly, we comment that, in the context of free-fermion problems, the notion of fragile topological phases is related to the idea of stable equivalence in the theory of vector bundles (most notably, in K-theory) \cite{Kitaev, MooreFreed, Read2017, fragile, ArisNoGo}. Consider a set of energy bands which is isolated from above and below by a continuous energy gap everywhere in the Brillouin zone. We say this set of band is trivial if one can find a full set of symmetric, localized Wannier functions. The classification of phases in this context then boils down to the study of equivalence of vector bundles under smooth deformation, which we will denote by the symbol ``$\sim$'' in this subsection.
For instance, consider two systems with respective valence-band vector bundles $\mathcal V_1$ and $\mathcal V_2$. We say they are in the same phase if $\mathcal V_1 \sim \mathcal V_2$.

However, in a K-theory-based classification a weaker sense of equivalence, termed ``stable equivalence,'' is required: even if $\mathcal V_1 \not \sim \mathcal V_2$, we say the two systems are stably equivalent so long as one can find a trivial bundle $\mathcal A$ such that $\mathcal V_1 \oplus \mathcal A \sim \mathcal V_2\oplus \mathcal A $. 
By our definitions, if a set of filled bands $\mathcal T$ is topological, one cannot find any trivial $\mathcal A$ for which $\mathcal T \sim \mathcal A$. Yet, in the spirit of stable equivalence, one should further examine the stability of the topological obstruction upon the addition of additional filled bands. We say $\mathcal T$ is fragile topological if one can find a pair of trivial bands $\mathcal B$ and $\mathcal C$ such that $\mathcal T \oplus \mathcal B \sim \mathcal C$; if no such pair can be found, we say $\mathcal T$ is stably topological.

We emphasize therefore, that, for example, the Hopf insulator \cite{Liu_1612} is \emph{not} an example of a fragile topological phase. The filled bands for this model correspond to a completely trivial vector bundle, and therefore it can be trivialized simply by adding unfilled bands. We believe that spatial symmetries are essential to obtain fragile topological phases.

\section{The general approach for establishing fragile topology}
\label{sec:general_approach}
\begin{figure}
\includegraphics[width=8cm]{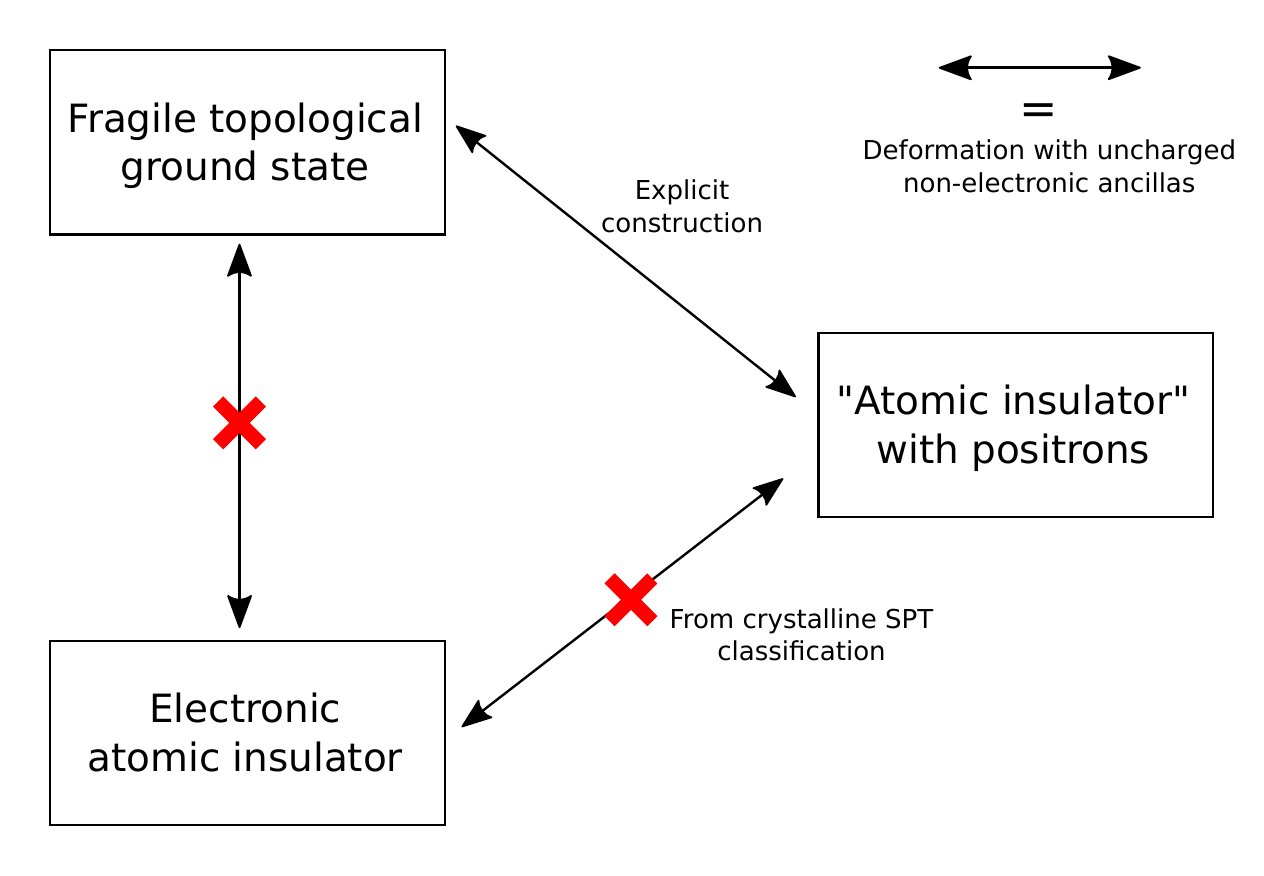}
\caption{\label{howto}How to establish that a given ground state is in a fragile topological phase.}
\end{figure}
In this section we briefly describe the general idea for how to establish that a ground state is in an interacting fragile topological phase. The specifics will be gone over in great detail for a specific example in Section \ref{sec:Square}. The components of the argument can be summarized in Figure \ref{howto}.

The idea is that we want to exploit the known classifications of crystalline SPT phases from Refs.~ \onlinecite{Song_1604,Thorngren_1612,Huang_1705,ElsePreparation}. As mentioned above, these classifications are based on an unphysical equivalence relation where non-electronic ancillas are allowed. By contrast, what we want to prove to establish a non-trivial fragile topological phase is that a state can never be deformed to an atomic insulator with purely electronic ancillas. But any such atomic insulator will obviously be a purely electronic one, so if we can prove that our initial ground state cannot be deformed into an electronic atomic insulator, \emph{even with} the addition of non-electronic ancillas as an intermediate step, then this an even stronger result, and the one we want follows.

To prove this result, the first step is to identify what crystalline SPT phase the ground state is in with respect to the classification of Refs.~\onlinecite{Song_1604,Thorngren_1612,Huang_1705,ElsePreparation}. The states we consider are always ``trivial'' with respect to this classification, which is to say that they can be deformed to an ``atomic insulator'' (possibly containing positrons). Since there can still be distinct trivial phases in the classification, we then use the ideas from Refs.~\onlinecite{Song_1604,Thorngren_1612,Huang_1705,ElsePreparation}  to determine whether the resulting atomic insulator can ever be deformed to a purely electronic atomic insulator.

\section{Decorated square lattice model
\label{sec:Square}}
In this section, we will first analyze in detail a fermionic model defined on a decorated square lattice, and from this present a general argument concerning the stability of certain fragile topological phases against interactions.

\subsection{Tight-binding model}
\label{subsec:TBM}
\begin{figure}
\begin{center}
\includegraphics[width=0.99\columnwidth]{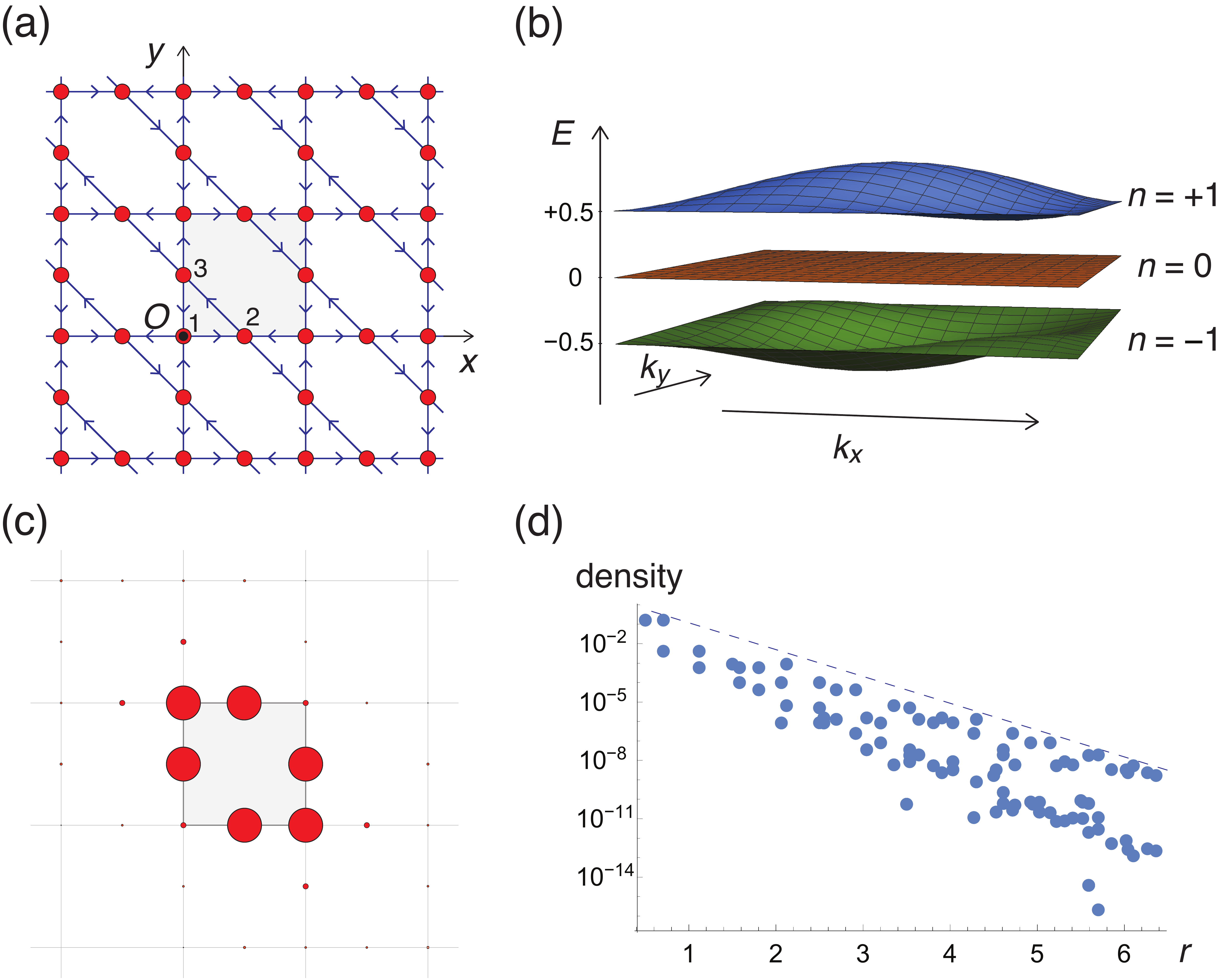}
\caption{\label{kagome1}(a) The tight-binding model on the decorated square lattice. The gray square is the unit cell that contains three sites $i=1,2,3$. (b) The band dispersion $E=\epsilon_{n,\vec{k}}$ of $\hat{H}_{\vec{k}}^{(1)}$, where $n=0, \pm1$ is the band index. (c) The real space plot of the Wannier orbital of the $n=0$ band. The size of each circle represents $|w_i(\vec{R})|$. (d) The exponential decay of the probability density $|w_i(\vec{R})|^2$ as a function of the distance $r$ from the Wannier center $\vec{x}=(\frac{1}{2},\frac{1}{2})$.}
\end{center}
\end{figure}

We begin by considering a model for a fractional topological phase in a free-fermion system, i.e.\ a \emph{fragile topological insulator} (FTI).
We start from the tight-binding model illustrated in Fig.~\ref{kagome1}(a). The model has three $s$-orbitals per unit cell, labeled by $i=1,2,3$ in Fig.~\ref{kagome1}(a), and is inversion symmetric. This lattice can actually be viewed as the kagome lattice. The kagome lattice model has been studied in detail in Refs.~\cite{Kagome,ArisNoGo} and we will use the results.  However, the minimum symmetry setting for our discussion requires only the inversion and the lattice translation, and here we regard the lattice as a decorated square lattice.  We set the lattice constant to be 1. 

The tight-binding Hamiltonian in the Fourier space reads
\begin{eqnarray}
\hat{H}_{\vec{k}}^{(1)}&=&t(1+e^{ik_x})\hat{c}_{2,\vec{k}}^\dagger\hat{c}_{1,\vec{k}}+t(1+e^{ik_y-ik_x})\hat{c}_{3,\vec{k}}^\dagger\hat{c}_{2,\vec{k}}\notag\\
&&\quad+t(1+e^{-ik_y})\hat{c}_{1,\vec{k}}^\dagger\hat{c}_{3,\vec{k}}+\text{h.c.}.
\end{eqnarray}
Here, $\hat{c}_{i,\vec{R}}^\dagger$ is the creation operator of a spinless electron on the site $i$ belonging to the unit cell $\vec{R}\in\mathbb{Z}^2$ and $\hat{c}_{i,\vec{k}}^\dagger$ is its Fourier transformation. The inversion symmetry $\hat{I}$ is implemented as $\hat{I}\hat{c}_{i,\vec{k}}^\dagger\hat{I}^{\dagger}=\rho_{i,\vec{k}}\hat{c}_{i,-\vec{k}}^\dagger$ with $\rho_{1,\vec{k}}=1$, $\rho_{2,\vec{k}}=e^{-ik_x}$, and $\rho_{3,\vec{k}}=e^{-ik_y}$ so that $\hat{I}\hat{H}_{\vec{k}}^{(1)}=\hat{H}_{-\vec{k}}^{(1)}\hat{I}$. We diagonalize $\hat{H}_{\vec{k}}^{(1)}$ and write 
\begin{equation}
\hat{H}_{\vec{k}}^{(1)}=\sum_{n=0,\pm1}\epsilon_{n,\vec{k}}\hat{\gamma}_{n,\vec{k}}^\dagger\hat{\gamma}_{n,\vec{k}}
\end{equation}
using the the creation operator $\hat{\gamma}_{n,\vec{k}}^\dagger=\sum_{i=1}^3(\vec{u}_{n,\vec{k}})_i\hat{c}_{i,\vec{k}}^\dagger$ of the Bloch state.   The dispersion relation $E=\epsilon_{n,\vec{k}}$ for $t=\frac{i}{4}$ is plotted in Fig.~\ref{kagome1}(b).  
The hopping parameter is chosen in such a way that $|\epsilon_{n,\vec{k}}|<1$ and that the magnitude of the band gap is $\sim0.5$.
The band index $n=0, \pm1$ coincides with the Chern number of each band, i.e., $C_n=\int\frac{d^2k}{2\pi i}\partial_{k_y}\vec{u}_{n,\vec{k}}\cdot\partial_{k_x}\vec{u}_{n,\vec{k}}+\text{c.c.}=n$~\cite{Kagome}.

The flat band ($n=0$) has a zero Chern number and thus admits a full set of symmetric, exponentially localized Wannier orbital by itself~\cite{ArisNoGo}.  Using the explicit form of the Bloch function
\begin{eqnarray}
\vec{u}_{0,\vec{k}}&=&(\tfrac{1+e^{ik_x-ik_y}}{1+e^{ik_x}},\tfrac{1+e^{-ik_y}}{1+e^{-ik_x}},1)^T\mathcal{N}_{0,\vec{k}},\label{Bloch0}\\
\mathcal{N}_{0,\vec{k}}&=&e^{-ik_x/2}\sqrt{\tfrac{1+\cos k_x}{3+\cos k_x+\cos k_y+\cos(k_x-k_y)}},
\end{eqnarray}
we can readily construct the Wannier function by the Fourier transformation, $w_{i}(\vec{R})=\int\frac{d^2k}{(2\pi)^2}e^{i\vec{k}\cdot\vec{R}}(\vec{u}_{0,\vec{k}})_i$.  As shown in Fig.~\ref{kagome1}(c), the Wannier center coincides with the plaquette center $\vec{x}=(\frac{1}{2},\frac{1}{2})$  and $|w_{i}(\vec{R})|^2$ decays exponentially with the distance $r$ from the Wannier center as demonstrated in Fig.~\ref{kagome1}(d).   This band gives rise to an obstructed phase as we discuss in Sec.~\ref{subsec:sqobstructed}.

\subsection{Fragile topological insulator}
\label{subsec:FTI}
\begin{figure}
\begin{center}
\includegraphics[width=0.99\columnwidth]{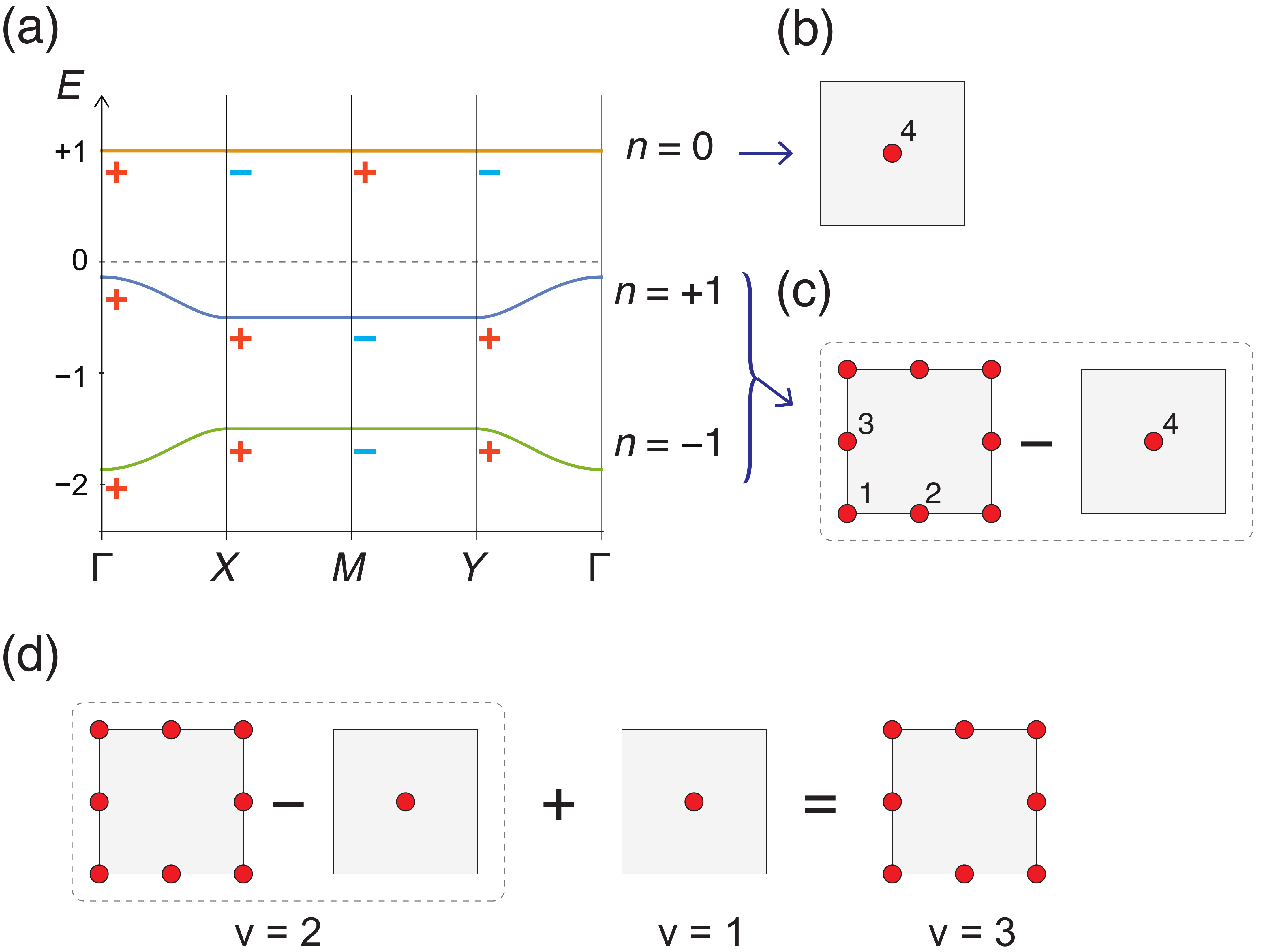}
\caption{\label{kagome2}(a) The band structure of $\hat{H}_{\vec{k}}^{(1)}+\hat{H}_{\vec{k}}^{(2)}$ along lines connecting TRIMs $\Gamma=(0,0)$, $X=(\pi,0)$, $Y=(0,\pi)$, $M=(\pi,\pi)$. (b) Illustration of an atomic insulator with an $s$-orbital sitting at the plaquette center. (c) Intuitive illustration of the FTI $|\Psi\rangle$. (d) A process of trivializing $|\Psi\rangle$ by stacking an atomic insulator of electrons. }
\end{center}
\end{figure}

Next, in order to realize a FTI, we shift the energy levels by adding
\begin{eqnarray}
\hat{H}_{\vec{k}}^{(2)}=-\sum_{i=1}^3\hat{c}_{i,\vec{k}}^\dagger\hat{c}_{i,\vec{k}}+2\hat{\gamma}_{0,\vec{k}}^\dagger\hat{\gamma}_{0,\vec{k}}
\end{eqnarray}
to $\hat{H}_{\vec{k}}^{(1)}$, so that the $n=\pm1$ bands sit below $E=0$ and the $n=0$ band is above $E=0$ [see Fig.~\ref{kagome2}(a)]. The first term of $\hat{H}_{\vec{k}}^{(2)}$ is just the on-site potential, and the second term can also be realized by exponentially-decaying hopping since the Bloch function $\vec{u}_{0,\vec{k}}$ in Eq.~\eqref{Bloch0} is analytic in $\vec{k}$. If desired, one can truncate the hopping at some finite range so that $\hat{H}_{\vec{k}}^{(2)}$ becomes strictly local.

Let $|\Psi\rangle$ be the state obtained by occupying the two bands below the chemical potential $\mu=0$. This is the unique gapped ground state of $\int\frac{d^2k}{(2\pi)^2}(\hat{H}_{\vec{k}}^{(1)}+\hat{H}_{\vec{k}}^{(2)})$ and it has the filling $\nu=2$.  The net Chern number is canceled out and a priori it could be smoothly deformable to a product state. However, the combination of inversion parities forms the obstruction to such a deformation within the single-particle problem of electrons.  In Fig.~\ref{kagome2}(a), we show the inversion parity of each band at the four time-reversal invariant momenta (TRIMs).  The two occupied bands have in total two odd parities at $\vec{k}=(\pi,\pi) $ and two even parities at other three TRIMs.  This combination of the inversion eigenvalue cannot be realized as a stacking of atomic insulators and thus cannot be topologically trivial. In fact, this parity combination implies a nontrivial flow of the Wilson loop eigenvalues~\cite{ArisPRB2014} and $|\Psi\rangle$ was identified as a noninteracting FTI in Ref.~\cite{ArisNoGo}.

In contrast, if an additional site is added at the plaquette center (which we call $i=4$ from now) for every unit cell, the $n=0$ band can be smoothly deformed into the product state illustrated in Fig.~\ref{kagome2}(b), as we explicitly demonstrate in Sec.~\ref{subsec:sqobstructed}.  This, in turn, implies that the insulator $|\Psi\rangle$ can be intuitively depicted as Fig.~\ref{kagome2}(c), in which the unfilled atomic limit in Fig.~\ref{kagome2}(b) is \emph{formally subtracted} from the product state of electrons occupying all three sites $i=1,2,3$.  In other words, the insulator $|\Psi\rangle$ can be trivialized by adding to it a product state (ancillas with $\nu=1$) in Fig.~\ref{kagome2}(d). This type of `fragile' topology was first proposed in Ref.~\onlinecite{fragile}. However, this requires changing the filling from $\nu=2$ to $\nu=3$.  Adding an unfilled band of electrons does not change the filling but cannot trivialize the fragile topology since, by definition, it does not change the representation of occupied bands.  

\subsection{Trivializing FTI by positrons}
\begin{figure}
\begin{center}
\includegraphics[width=0.99\columnwidth]{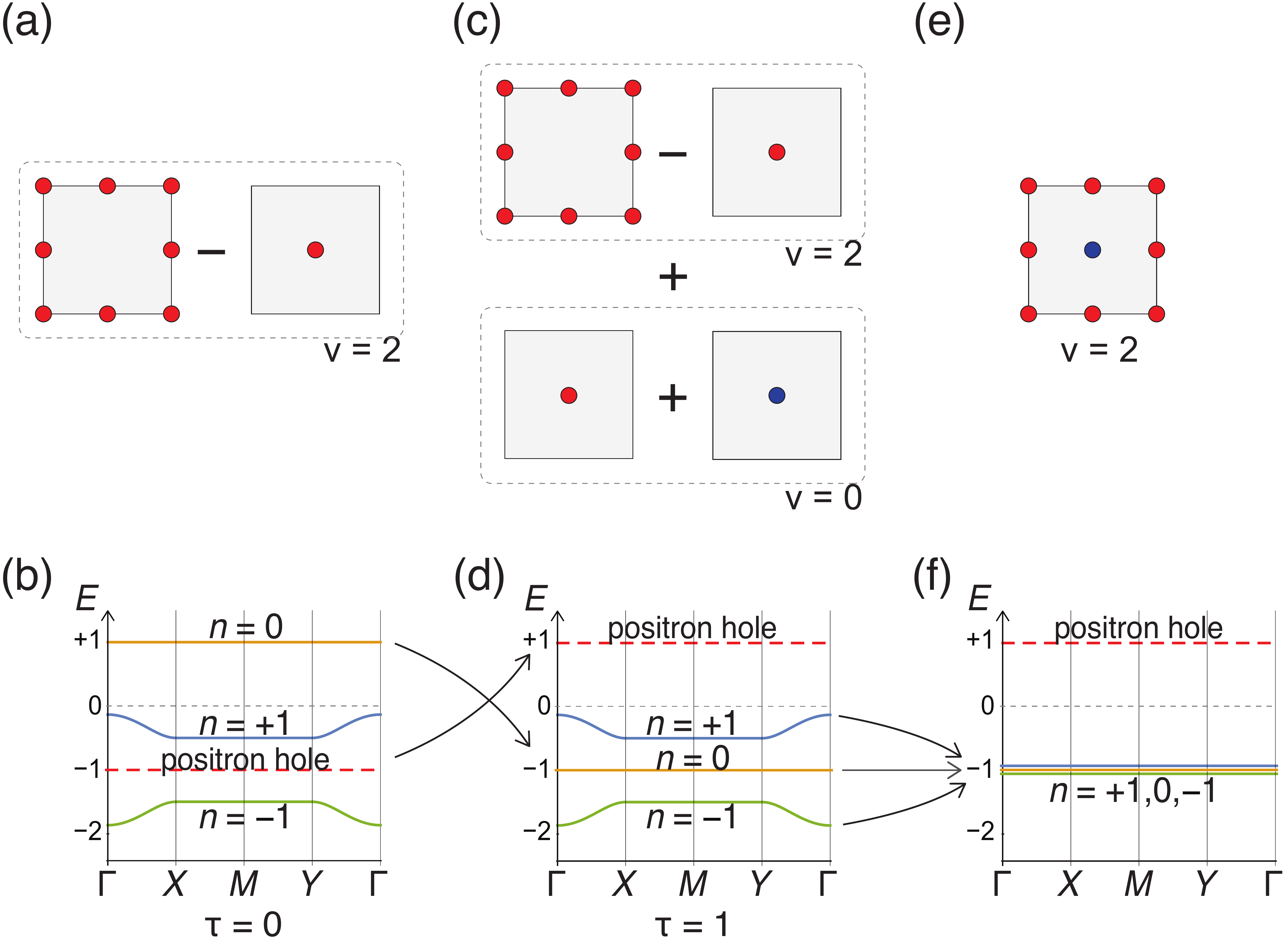}
\caption{\label{kagome3}
(a) Illustration of $|\Psi\rangle$, the ground state of $\int\frac{d^2k}{(2\pi)^2}\hat{H}_{\vec{k}}(\tau)$ at $\tau=0$.
(b) The band structure of  $\hat{H}_{\vec{k}}(\tau)$ at $\tau=0$.
(c) Illustration of the ground state of $\int\frac{d^2k}{(2\pi)^2}\hat{H}_{\vec{k}}(\tau)$ at $\tau=1$, in which the $n=0$ electronic band and the positron band are occupied. 
The red and blue circles respectively represent electrons and positrons in an $s$-orbital.
(d) The band structure of  $\hat{H}_{\vec{k}}(\tau)$ at $\tau=1$.
(e) The product state limit of (c). (f) The band structure of  $\hat{H}_{\vec{k}}(\tau)$ at $\tau=1$ after switching off the hopping in $\hat{H}_{\vec{k}}^{(1)}$.}
\end{center}
\end{figure}

So far, as in past works, the ``subtraction'' shown in Fig.~\ref{kagome2}(c) has been treated as a purely formal one. However, we now take a step towards interpreting it more concretely: ``minus an electron'' is heuristically the same as a \emph{positron}, i.e. a particle carrying charge $-e$. In this section, we propose to take this picture seriously, and use it to understand properties of the FTI. In particular, we will explicitly demonstrate that, by adding extra degrees of freedom such that the Hilbert space also contains positronic states (in the language of Section \ref{sec:ancillas}, adding a non-electronic ancilla), the FTI $|\Psi\rangle$ in Fig.~\ref{kagome3}(a) can be smoothly deformed into a product state of electrons and positron as illustrated in Fig.~\ref{kagome3}(e), where red and blue circles respectively represent electrons and positrons in an $s$-orbital. This deformation can take place purely in the space of free-fermion Hamiltonians.

Let us make an immediate remark. By adding positronic degrees of freedom we are violating the condition on the deformations that we imposed in defining fragile topology summarized in Sec.~\ref{sec:definitions}. So the existence of the deformation we are constructing does not contradict with our earlier statements on fragile topology. Thinking about it in a different way, since we will initially be adding an \emph{empty} band of positrons (which does not change the filling), 
this example also serves to illustrate the necessity of forbidding the presence of opposite-signed charge states in the Hilbert space when defining fragile topology, which might not have been immediately obvious.

To this end, let $\hat{p}_{\vec{R}}$ be the creation operator of a positron at the plaquette center (the site $i=4$) of the unit cell $\vec{R}$. 
We introduce an \emph{empty} band of positrons by adding a positive on-site potential term $+\sum_{\vec{R}}\hat{p}_{\vec{R}}^\dagger\hat{p}_{\vec{R}}$. In terms of the the hole operator of positron $\hat{h}_{\vec{R}}\equiv\hat{p}_{\vec{R}}^\dagger$, this corresponds to the term
\begin{eqnarray}
\hat{H}_{\vec{k}}^{(3)}=-\hat{h}_{\vec{k}}^\dagger\hat{h}_{\vec{k}}
\end{eqnarray}
in the momentum space, giving rise to a flat band at $E=-1$ [the red dashed line in Fig.~\ref{kagome3}(b)]. The U(1) charge operator $\hat{Q}$ is now given by $\hat{Q}=\sum_{\vec{R}}\left(\sum_{i=1}^3\hat{c}_{i,\vec{R}}^\dagger \hat{c}_{i,\vec{R}}+\hat{h}_{\vec{R}}^\dagger\hat{h}_{\vec{R}}-1\right)$ and the inversion $\hat{I}$ acts on $\hat{h}_{\vec{k}}$ as $\hat{I}\hat{h}_{\vec{k}}^\dagger\hat{I}^{\dagger}=e^{-i(k_x+k_y)}\hat{h}_{-\vec{k}}^\dagger$.  We also include the mixing term of the $n=0$ band and the positron hole band:
\begin{eqnarray}
\hat{H}_{\vec{k}}^{(4)}(\tau)&=&(1-\cos\tau\pi)(\hat{h}_{\vec{k}}^\dagger\hat{h}_{\vec{k}}-\hat{\gamma}_{0,\vec{k}}^\dagger\hat{\gamma}_{0,\vec{k}})\notag\\
&&+\sin\tau\pi (\hat{\gamma}_{0,\vec{k}}^\dagger\hat{h}_{\vec{k}}+\hat{h}_{\vec{k}}^\dagger\hat{\gamma}_{0,\vec{k}}),
\end{eqnarray}
and the total Hamiltonian reads
\begin{equation}
\hat{H}_{\vec{k}}(\tau)=\sum_{\ell=1}^3\hat{H}_{\vec{k}}^{(\ell)}+\hat{H}_{\vec{k}}^{(4)}(\tau).
\end{equation}

When $\tau=0$, the mixing term $\hat{H}_{\vec{k}}^{(4)}(0)$ vanishes and the positron hole band locates at $E=-1$ (i.e., the positron states are not occupied) while the unoccupied $n=0$ electron band is at $E=+1$ [Fig.~\ref{kagome3}(a,b)]. The two flat bands interchange as $\tau$ increases \emph{without closing the band gap}.  Namely, the wavefunction of the two flat bands mixes and the weight changes but the energy levels stay unchanged. When $\tau$ reaches $1$, the $n=0$ electron band has $E=-1$ and the positron hole band has $E=+1$, meaning that the added positron state as well as the $n=0$ electron band are fully occupied [Fig.~\ref{kagome3}(c,d)]. The Hamiltonian at $\tau=1$ has the simple form
\begin{eqnarray}
\hat{H}_{\vec{k}}(\tau = 1)=\hat{H}_{\vec{k}}^{(1)} +\hat{h}_{\vec{k}}^\dagger\hat{h}_{\vec{k}} -\sum_{i=1}^3\hat{c}_{i,\vec{k}}^\dagger\hat{c}_{i,\vec{k}}.
\end{eqnarray}
This is simply the original tight-binding model with on-site potentials. Finally, one can smoothly switch off the hopping in $\hat{H}_{\vec{k}}^{(1)}$ and make all bands completely trivial as in Fig.~\ref{kagome3}(e,f). This completes the adiabatic process of deforming the FTI in (a) into the product state illustrated in (e).

\begin{figure}
\begin{center}
\includegraphics[width=0.99\columnwidth]{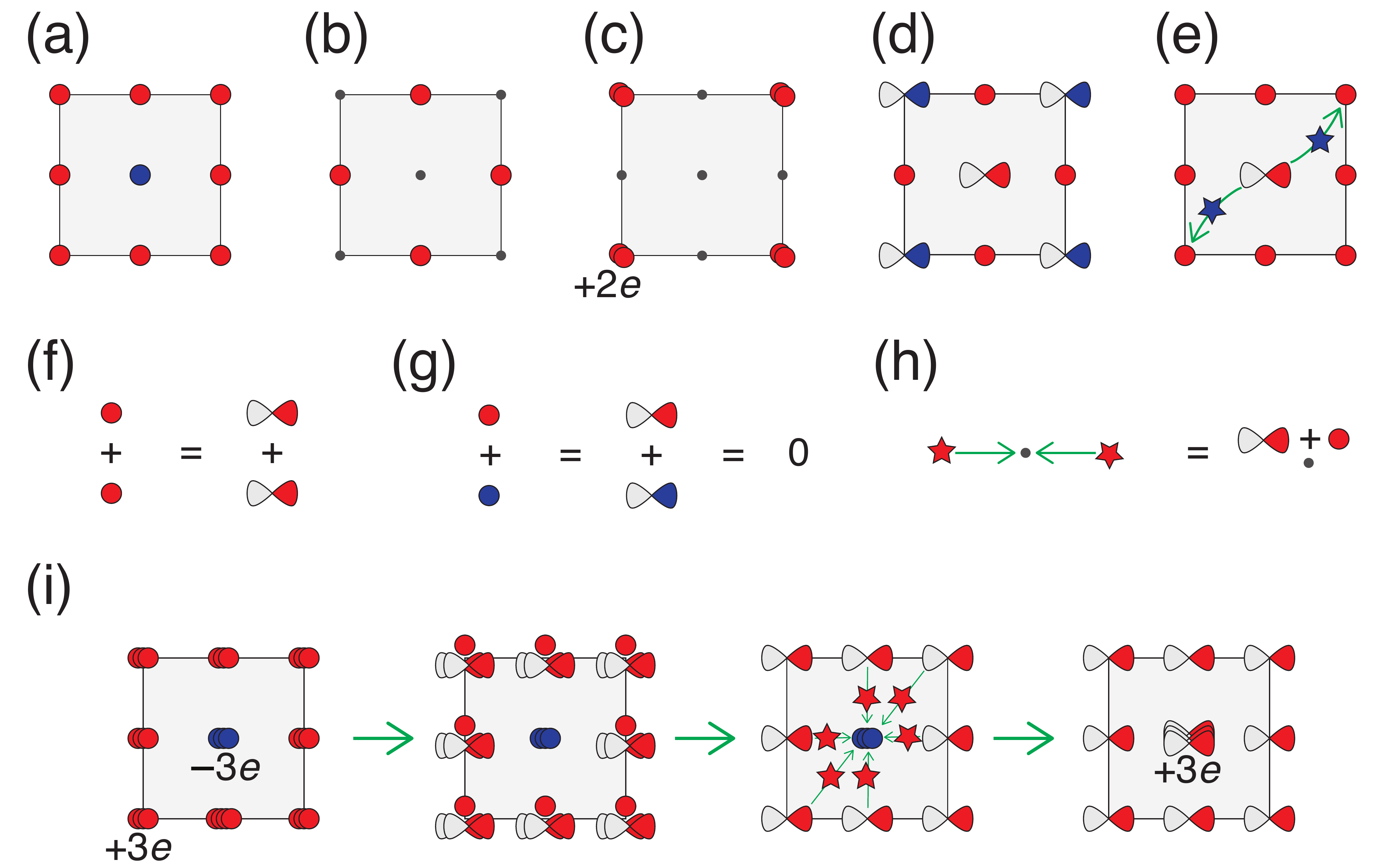}
\caption{\label{kagome4}(a)--(e): Illustration of product states of electrons and positrons, where 
$s$-($p$-)orbitals are shown by circles (two fins). Small black dots are the center of an inversion.  (f),(g): The identification rules among the product states, assuming interactions. (h): The symmetric deformation process of fermionic product states. Orbitals at nonsymmetric sites are illustrated by stars. (i) Three copies of (a) can be smoothly deformed into a product state of electrons only.}
\end{center}
\end{figure}

\subsection{Lattice homotopy of atomic insulators}
\label{sec_lattice_homotopy}
Here we show that the product state of positrons and electrons in Fig.~\ref{kagome4}(a) cannot be smoothly deformed into any product state of only electrons with the same filling, such as the ones in (b) and (c). This statement will be true even with ancillas. In this section, we allow non-electronic ancillas -- after all, we already introduced them in constructing the deformations in the previous section. We do, however, still demand that the ancillas be uncharged (recall the definitions of ``uncharged'' and ``electronic'' from Section \ref{sec:ancillas}). Thus, we are now considering the equivalence relation implicit in Refs.~\onlinecite{Song_1604,Thorngren_1612,Huang_1705,ElsePreparation}.

The first step is to show that that the deformation cannot take place though a continuous path in the space of atomic insulators -- an ``atomic-insulator homotopy''. In this section, by ``atomic insulator'' we just mean a product state; each site could by occupied by electrons or positrons, or both.

At each inversion symmetric point, filled orbitals can be classified into $s$- or $p$-types, depending on their inversion parity $\pm1$. In Fig.~\ref{kagome4}, filled $s$-orbitals are shown by circles and $p$-orbitals are illustrated by two fins.  One should understand these figures under the identification rules listed in (f) and (g), which are derived assuming the presence of interactions. For example, (f) says the state with two filled $s$-orbitals cannot be distinguished from the state with two filled $p$-orbitals 
Under these rules, the product states at a single inversion symmetric point are classified as $\mathbb{Z}\times\mathbb{Z}_2$, respectively corresponding to the U(1) charge and the inversion parity. These identification rules come from allowing strong interactions; for non-interacting systems, two filled $p$-orbitals are different from two filled $s$-orbitals, but with interactions we can do a continuous deformation acting that transforms one into the other. Specifically, we have the continuous path of inversion-symmetric on-site states $\ket{\phi_\theta} = \cos \theta \ket{s,s} + \sin \theta \ket{p,p}$, $\theta \in [0,\pi/2]$.

When the translation symmetry is included, there are four inversion symmetric sites ($i=1,\ldots,4$) in a unit cell.  A process of symmetric deformation of product states is summarized in the panel (h): a pair of inversion-related charges can be smoothly moved to the inversion center and then they split into an $s$ and a $p$-orbital.  
 Now, recall that the classification about a single inversion center is given by $\mathbb Z \times \mathbb Z_2$, where the $\mathbb Z$ factor indicates the charge localized at the point, and $\mathbb Z_2 = \{ -1,+1\}$ encodes the parity of the state. 
Under the deformation described above, the state is modified by the filling of an additional pair of orbitals of the $s$ and $p$ characters, respectively. This increases the charge by $2$, and at the same time flips the parity of the state. In more mathematical terms, by identifying configurations which are identical under atomic-insulator homotopy, the classification is reduced to $\mathbb Z \times \mathbb Z_2/ \langle (2,-1) \rangle \simeq \mathbb Z_4$. 

While this discussion is applicable to any of the four inversion centers in the unit cell, the total charge in a unit cell is also conserved in the presence of translation symmetry.
Furthermore, one can define a ``total parity'' that is also invariant under atomic-insulator homotopy: suppose we first deform the ground state to a strictly localized limit $\otimes_{c} | c\rangle$, where $c$ runs over all the inversion centers in the system.
Let $\hat P_c$ be the parity operator about $c$, and we define the local parity $\xi_c$ as the eigenvalue $\hat P_c |c\rangle = \xi_c | c \rangle$. While the sign of a single $\xi_c$ is ambiguous, the product 
$$\xi \equiv \prod_{c \text{ in a unit cell}} \xi_c$$
 is an unambiguous invariant, as the symmetric deformation requires the sending of pairs of charges from one inversion center $c_0$ to an inequivalent one $c_1$ in the unit cell, thereby keeping the product $\xi_{c_0} \xi_{c_1}$ unchanged.

If we fix the total charge and total parity in the unit cell, then knowing the $\mathbb{Z}_4$ invariant associated with three of the inversion symmetric points is enough to know it at the fourth point. Therefore, we obtain a $\mathbb{Z}\times\mathbb{Z}_2\times(\mathbb{Z}_4)^3$ classification of atomic insulators, where the $\mathbb{Z}$ factor represents the filling $\nu$, the $\mathbb{Z}_2$-factor corresponds to the mentioned total parity, and the remaining $\mathbb Z_4$ factors correspond to the three independent $\mathbb{Z}_4$ invariants associated with the inversion symmetric points. More explicitly, the three $\mathbb{Z}_4$-factors can be 
generated by the neutral configuration with one electron at the origin ($i=1$) and one positron at one of the other three symmetric sites $i\neq1$.~\footnote{Although our discussion here is for the inversion symmetry satisfying $\hat{I}^2=+1$, the classification results are unchanged even for $\hat{I}^2=(-1)^{\hat{Q}}$ (the fermion parity), which is naturally realized by the two-fold rotation for spinful electrons.}

The product state in Fig.~\ref{kagome4}(a) can be adiabatically deformed into, for example, the one in (d) through the process of dragging a pair of negative charges in (e). 
However, it cannot be deformed into a product state of only electrons. To see the obstruction, note that the total U(1) charge in each inversion center must be preserved modulo two in any symmetric deformation process. At filling $\nu=2$, a product state of electrons must contain at least two vacant inversion centers (i.e., even U(1) charge) as shown in (b) and (c). Therefore, the product state in the panel (a), where every inversion center has an odd U(1) charge, cannot be smoothly deformed into such states.

Now we must ask whether two states can be related by a local unitary even if they are not related by a a lattice homotopy. An argument that they cannot relies on the ``block state'' picture of crystalline topological phases introduced in Refs.~\onlinecite{Song_1604,Huang_1705}. This framework can also be derived~\cite{ElsePreparation} from the general approach of Ref.~\onlinecite{Thorngren_1612}. In the framework of Refs.~\onlinecite{Song_1604,Huang_1705,ElsePreparation}, an atomic insulator is called a block state of block dimension $0$. Moreover, the space of local relating two block states is captured by homotopies of block states of dimension $0, 1, \cdots, d-1$, where $d$ is the space dimension. We have already considered homotopies of block states of dimension $0$ (atomic insulator homotopies). Moreover, in the present context $d=2$, and there are no block states of dimension 1 because for any non-trivial 1-dimensional submanifold of space, the effective symmetry group (the subgroup of the full symmetry group that leaves the submanifold invariant) is $\mathrm{U}(1)$, and there are no non-trivial 1-dimensional fermionic symmetry-protected topological (SPT) phases with symmetry $\mathrm{U}(1)$.

These considerations, however, are rather abstract and depend on the validity of the classification of Ref.~\onlinecite{Thorngren_1612}, which has not been proven rigorously. In fact, in the present context we can give a more straightforward and rigorous argument. The important point is that we did not really need to use translation symmetry in the above argument. To show that the state shown in Fig.~\ref{kagome4}(a) is distinct from any atomic insulator of electrons, it is sufficient to show if two atomic insulators are related by a local unitary that is symmetric with respect to $\mathrm{U}(1)$ and inversion symmetry about one particular point, then they are related by a lattice homotopy with respect to those same symmetries. This can be shown using coarse-graining, as shown in Appendix \ref{appendix_equivalencebeyondhomotopy}. To see why this is sufficient, note the argument that Fig.~\ref{kagome4}(a) cannot be deformed to an electronic atomic insulator only relied on the fact that the charge at the inversion center is invariant modulo 2 (which follows just from inversion symmetry) and the invariance of the filling.

 Having completed the program layed out in Section \ref{sec:general_approach}, we conclude that the decorated square lattice model is indeed in a fragile topological phase that is stable to interactions.

\subsection{Obstructed topological phase}
\label{subsec:sqobstructed}

\begin{figure}
\begin{center}
\includegraphics[width=0.79\columnwidth]{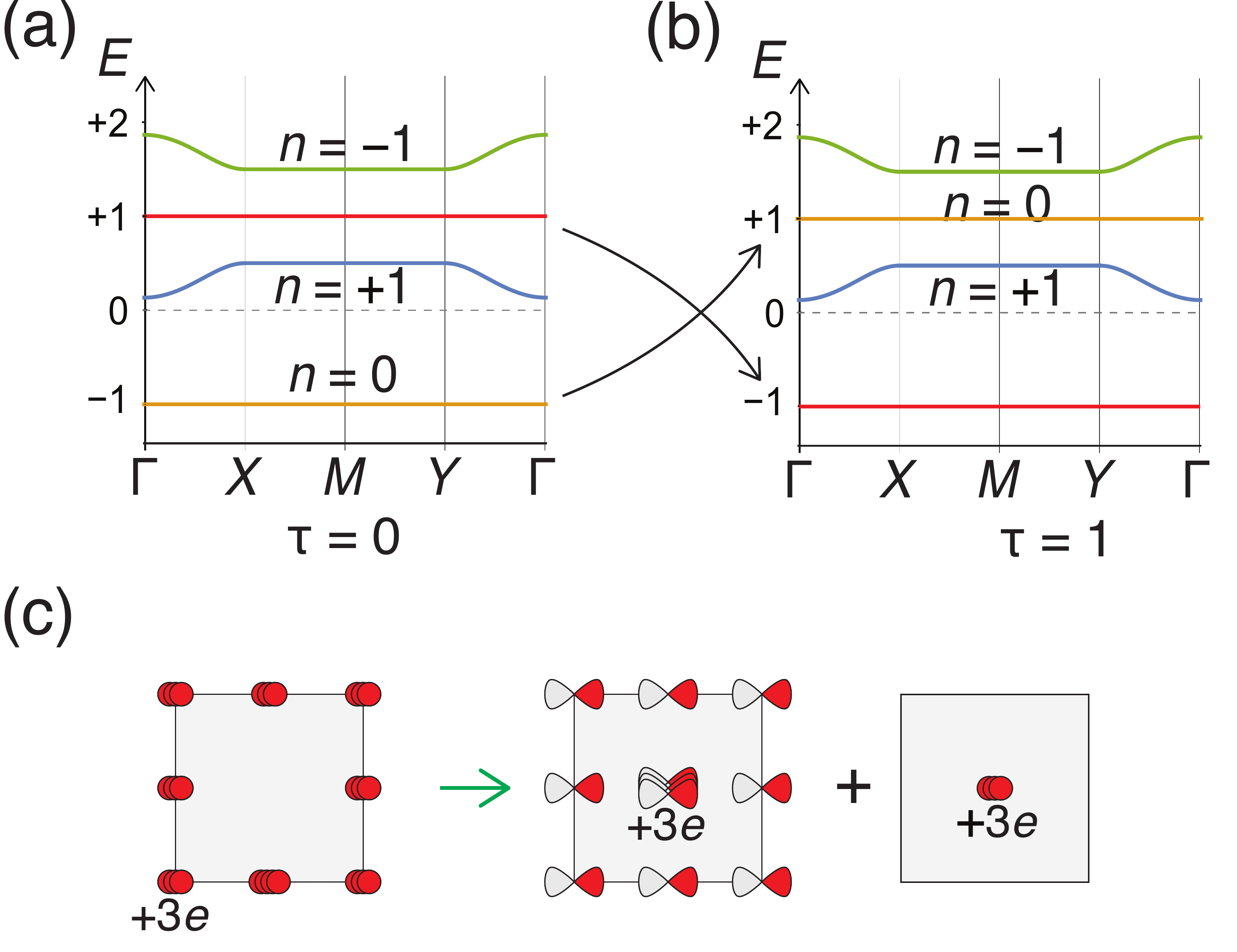}
\caption{\label{kagome5}
(a),(b) The band structure of $-\hat{H}_{\vec{k}}(\tau)$ at $\tau=0$ and $1$. The red line corresponds to the atomic insulator, illustrated in (d), with an $s$-orbital sitting at the plaquette center.
(c) With interactions, three copies of the original tight-binding orbitals can be moved in such way that trivialize the obstructed trivial insulator through a similar process as in Fig.~\ref{kagome4}(h).}
\end{center}
\end{figure}

We have shown that the band insulator that fills the $n=\pm1$ bands is in a fragile topological phase. Here we discuss that its particle-hole conjugate that occupies only the $n=0$ band is obstructed.  

To this end, let us consider the inverted single-particle Hamiltonian $-\hat{H}_{\vec{k}}^{(1)}-\hat{H}_{\vec{k}}^{(2)}$. Let $|\overline{\Psi}\rangle$ be the filling $\nu=1$ band insulator that fully occupies the $n=0$ band.   $|\overline{\Psi}\rangle$ is obstructed because the $n=0$ band has a symmetric localized Wannier orbital centered at $\vec{x}=(\frac{1}{2},\frac{1}{2})$ as shown in Sec.~\ref{subsec:TBM}, but there is no atomic site at the Wannier center in this model.  Moreover, it is obstructed within the electronic problem, because one cannot move any sites in the original tight-binding model unless positrons are allowed.

Now we demonstrate that, once a new electronic site at the plaquette center is introduced,  $|\overline{\Psi}\rangle$ is smoothly deformed to a product state.  In fact, we can reuse the same interpolating Hamiltonian as above with only an additional minus sign:
\begin{equation}
-\hat{H}_{\vec{k}}(\tau)=-\sum_{\ell=1}^3\hat{H}_{\vec{k}}^{(\ell)}-\hat{H}_{\vec{k}}^{(4)}(\tau).
\end{equation}
This time, one should interpret $\hat{h}_{\vec{k}}$ in $\hat{H}_{\vec{k}}^{(3)}$ and $\hat{H}_{\vec{k}}^{(4)}(\tau)$ as the annihilation operator of an electron (not a hole of positron) associated with the site $i=4$. Hence, the U(1) charge operator is given by $\hat{Q}=\sum_{\vec{R}}\left(\sum_{i=1}^3\hat{c}_{i,\vec{R}}^\dagger \hat{c}_{i,\vec{R}}+\hat{h}_{\vec{R}}^\dagger\hat{h}_{\vec{R}}\right)$.

At $\tau=0$, the $n=0$ band has $E=-1$, while the flat band of the added product state has $E=+1$ and is unoccupied [Fig.~\ref{kagome5}(a)]. As $\tau$ increases, they mix and interchange without closing the band gap, as shown in the panel (b). At $\tau=1$, the occupied band below $E=0$ is precisely the product state of localized $s$-orbital at the $i=4$ site, illustrated in Fig.~\ref{kagome2}(b).

One can readily prove the stability of the obstructed atomic insulator using this product state limit.  There are only $3$ possible product states at filling $\nu=1$, obtained by localizing the charge in $s$-orbital to one of the sites $i =1,2,3$, in the original tight-binding model $-\hat{H}_{\vec{k}}^{(1)}-\hat{H}_{\vec{k}}^{(2)}$.   Clearly, the product state in Fig.~\ref{kagome2}(b) cannot be adiabatically deformed to any one of these candidate product states, since such a deformation would violate the conservation of U(1) charge modulo 2 at the plaquette center.

\subsection{Breakdown of noninteracting fragile/obstructed phases}
\label{sec_breakdown}
It has been shown in Ref.~\onlinecite{ArisPRB2014} that the band insulator composed of $N$-copies ($N=2,3,4,\ldots$) of the above FTI, which has $2N$ odd parities at $\vec{k}=(\pi,\pi)$ and $2N$ even parities at other TRIMs, is still topological within the single-particle problem.  Here we discuss the stability of such states against many-body interactions.

Once positrons are introduced, the $N$-copies of the FTI can be smoothly deformed to the $N$-copies of the product state of positrons and electrons in Fig.~\ref{kagome4}(a).  For example, the leftmost panel in Fig.~\ref{kagome4}(i) is the case of $N=3$.  When $N=2$, it can be readily seen that positrons in the product state cannot be eliminated in any deformation process. \footnote{The state can be deformed to a product state with only non-negative charges, but which still carries non-trivial eigenvalue of inversion at the plaquette center even though $q=0$ there. This, however, is not a purely electronic atomic insulator because the only purely electronic state with $q=0$ is the Fock vacuum, which carries trivial inversion eigenvalue.} 

 However, three copies of the FTI can be adiabatically deformed into a product state of electrons only, as illustrated in Fig.~\ref{kagome4}(i), through the symmetric deformation path discussed in Sec.~\ref{sec_lattice_homotopy}. This suggests that some fragile non-interacting topological phases may become trivial in the presence of interactions.  Note that the deformation process in (i) cannot be taken as a proof of this claim, since  one must find a path to an electron product state without introducing positrons in any intermediate step.

Similarly, the obstruction in the three copies of the obstructed atomic insulator can be resolved by smooth deformation of the tight-binding orbitals as illustrated in Fig.~\ref{kagome5}(c). This process does not involve any positrons but assume strong interactions among electrons. Thus, the three copies of the obstructed atomic insulator can be smoothly trivialized.

\section{Model with spinful electrons and time-reversal}
\label{sec:honeycomb}
Having established the interaction stability of the fragile topology for the model of spinless fermions in Sec.~\ref{sec:Square}, it is of physical interest to ask if the same stability is found for models of spinful electrons with time-reversal $\mathcal T$ symmetries, as this symmetry setting is naturally realized in crystalline materials with strong spin-orbit coupling and no magnetic ordering.

In fact, the fragile model first described in Ref.\ \onlinecite{fragile} concerns exactly with $\mathcal T$-invariant spinful electrons. In this subsection, we will show that this noninteracting FTI is also stable against the introduction of interaction. To establish this, we first note that we will only need to use the charge quantum number in our argument; while the other point-group symmetries could in general lead to additional quantum numbers in the interacting setting \cite{FragileMott}, these additional quantum numbers can be ignored for the following analysis
\footnote{
We remark that, in the special case of atomic insulators smoothly deformable into a non-interacting limit, the only nontrivial quantum number is the charge. 
}. 
Furthermore, due to Kramers degeneracy, the total charge localized to any site is quantized to $q n$ with $n\in \mathbb Z$ and $q \equiv 2 e$ is twice the charge of the spinful electron.

\begin{figure}
\includegraphics[width=0.99\columnwidth]{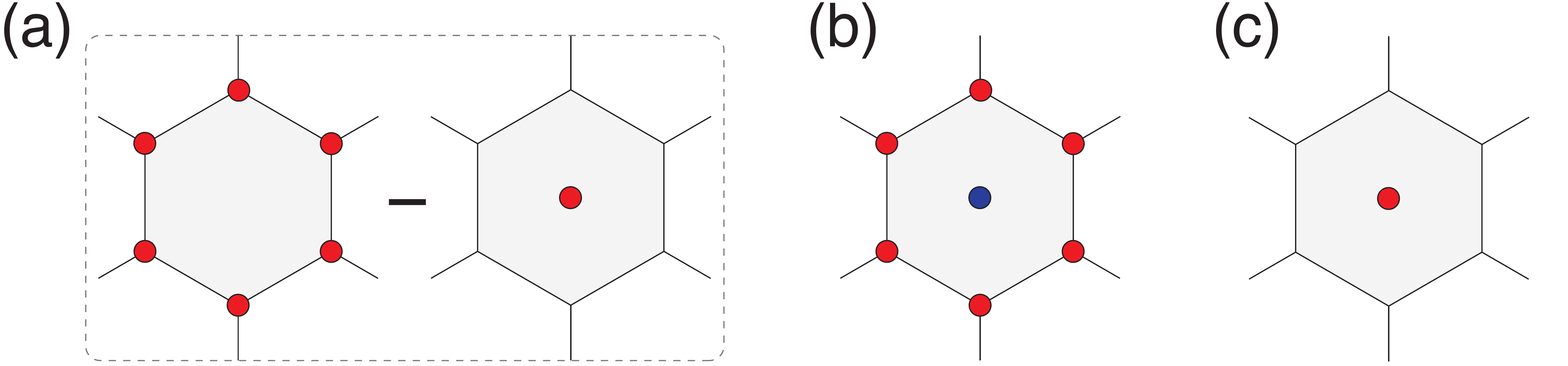}
\caption{\label{fig_honeycomb}
(a) Intuitive illustration of the honeycomb FTI.
(b) The product state limit of (a), which requires both electrons and positrons.
(c) The product state of only electrons at the same filling as (b).
}
\end{figure}

Let us study the stability of the model in Ref.\ \onlinecite{fragile}. The model there is defined on the honeycomb lattice, and the spatial symmetries are described by the wallpaper group 17 ($p6mm$)~\cite{ITC}, generated by the two lattice translations along $\vec a_{1,2}$, a six-fold rotation $C_6$ about the point-group origin (a plaquette center of the honeycomb lattice) in the 2D plane, and a mirror along (say) $\vec a_1$. 
Now, observe that the honeycomb FTI state is constructed by filling two bands, whose complement corresponds to an atomic insulator with a Kramers pair of electron localized to the plaquette center [Fig.~\ref{fig_honeycomb}(a)], which is in an obstructed atomic limit~\cite{Bradlyn17} unless new sites at the plaquette center is introduced.
Using the same construction as in Sec.~\ref{sec:Square}, by admitting positrons into the system one can deform the FTI into the product state shown in Fig.~\ref{fig_honeycomb}(b), where charge $-q$ is pinned to the origin, and charge $q$ is pinned to each of the honeycomb sites
\footnote{
If desired, one can also compute the atomic-insulator homotopy classification by a a similar computation as the one in Sec.~\ref{sec_lattice_homotopy}.
}.
Importantly, as the honeycomb site is invariant under a $C_3$ rotation, the local charge there must be conserved modulo $3q$ under homotopy. Similarly, the local charge at the origin is conserved modulo $6q$.  We thus conclude the point charges in Fig.~\ref{fig_honeycomb} are all immobile, implying the impossibility to eliminate the positrons and obtaining a purely electronic product state. This establishes the interaction stability of the FTI in Ref.~\onlinecite{fragile}.

\section{Bosonic model}
\label{sec:bosons}

The basic philosophy of how to establish fragile protection of interacting topological phases, discussed above for fermions, works equally well for bosons. (To avoid confusion, in this section we will use the terminology ``sign-restricted ancilla'' to refer to what we previously called an ``electronic ancilla'', though the definition of the latter in Section \ref{sec:ancillas} did not actually depend on the particles being fermionic.) Consider a bosonic system with $\mathrm{U}(1)$ symmetry. Then, for example, any ground state built from only positive charges, but which, by adding non-sign-restricted ancillas, can be smoothly deformed to the state shown in Fig.~\ref{fig_honeycomb}(b) without breaking charge conservation or the symmetries of the honeycomb lattice, must be fragile protected (in the presence of $\mathrm{U}(1)$ and the lattice symmetries), by similar arguments to those discussed above.

The only outstanding question is whether such a bosonic state exists. Here we will show that this is equivalent to a different question which has already been considered in the literature: the existence of a featureless Mott insulator on the honeycomb lattice with hard-core bosons hopping on the vertices, and site filling $1/2$ (that is, average charge $1$ per two-vertex unit cell). Here, ``featureless'' means that no symmetries are spontaneously broken and there is no fractionalization.
Numerical evidence for existence of such a state has been presented in Refs.~\onlinecite{Kimchi_1207,Ware_1507}. We refer to the state constructed in Refs.~\onlinecite{Kimchi_1207,Ware_1507} as the honeycomb Mott state $|\Psi_{\text{HM}}\rangle$.

First of all, we note that the state $|\Psi_{\text{HM}}\rangle$ as constructed in Ref.~\onlinecite{Kimchi_1207} is explicitly time-reversal invariant, but we consider phases not requiring time-reversal for protection; that is, we allow the time-reversal symmetry to be explicitly lifted.  We will also assume that the honeycomb Mott state is not protected by $\mathrm{U}(1)$ alone; this should be clear, because non-trivial $\mathrm{U}(1)$ SPTs are characterized by their Hall conductance, and the state $|\Psi_{\text{HM}}\rangle$ was explicitly constructed to be time-reversal invariant. (The fact that we then allowed time-reversal to be lifted does not matter, because even a time-reversal breaking perturbation cannot change the Hall conductance without a phase transition).

Now we can apply the general framework of Refs.~\onlinecite{Song_1604,Huang_1705,ElsePreparation} (see the  discussion of this framework in Sec.~\ref{sec_lattice_homotopy}), which shows that the state $|\Psi_{\text{HM}}\rangle$ should be symmetrically deformable (in the presence of uncharged, non-sign-restricted ancillas) to a ``block state'', as defined there.
Given the assumptions above, this block state cannot carry non-trivial SPTs on surfaces [on which the effective internal group is just $\mathrm{U}(1)$], nor on mirror lines [because the effective internal group is $\mathbb{Z}_2 \times \mathrm{U}(1)$, and $\mathcal{H}^2(\mathbb{Z}_2 \times \mathrm{U}(1), \mathrm{U}(1)) = 0$]. That is, the block state must be built of zero-dimensional blocks (points). Each such zero-dimensional block carries some integer charge under $\mathrm{U}(1)$. There are only two such charge distributions consistent with overall $1/2$ site filling: they are shown in Fig.~\ref{fig_honeycomb}(b) and (c). Note that the states shown in (b) and (c) are in \emph{different} phases with respect to the symmetries of the honeycomb lattice, because if we only keep the $C_6$ rotation symmetry about a given plaquette center, then (b) is deformable to a state with only $\mathrm{U}(1)$ charge $-1$ at the plaquette center, and the arguments of Appendix \ref{appendix_equivalencebeyondhomotopy} show that such a state cannot be deformed to a state with only charge $+1$ at the inversion center, because $-1 \neq 1 \, \mathrm{mod} \, 6$.

We might guess that the one into which $|\Psi_{\text{HM}}\rangle$ is deformable is the one shown in (c), because this roughly corresponds to the intuition used in the construction of Ref.~\onlinecite{Kimchi_1207}: inserting a boson in a superposition over the vertices of each plaquette. To see this is indeed the case, note that the invariant that distinguishes (b) and (c) -- the $\mathrm{U}(1)$ charge at the plaquette center, mod 6 -- can be detected as the \emph{total} charge of a finite (but large) system with boundary that is symmetric with respect to $C_6$ rotation about a plaquette center. The construction of Ref.~\onlinecite{Kimchi_1207} can be applied in such a system with boundary, and produces a state with charge 1 mod 6; hence, we conclude that the state $|\Psi_{\text{HM}}\rangle$ is deformable into the configuration shown in (c) once proper ancillas are introduced at the plaquette center. Hence, the honeycomb Mott state $|\Psi_{\text{HM}}\rangle$ by itself is in an obstructed topological phase.

Now let us consider the particle-hole conjugate of $|\Psi_{\text{HM}}\rangle$, which we call $|\overline{\Psi}_{\text{HM}}\rangle$. This is constructed from $|\Psi_{\text{HM}}\rangle$ by applying the charge-conjugation operator $\ket{0}\bra{1} + \ket{1}\bra{0}$ at each vertex (where $\ket{0}$ and $\ket{1}$ denote the unoccupied and singly-occupied states respectively). Note that this state is still $\U(1)$ symmetric, and has the same filling. By similar arguments to the above, we conclude that it must be deformable into either Fig.~\ref{fig_honeycomb}(b) or Fig.~\ref{fig_honeycomb}(c). Moreover, by considering the charge on a system with $C_6$-symmetric boundary, we see that it must be (b). Therefore, we conclude that the state $|\overline{\Psi}_{\text{HM}}\rangle$ is a fragile topological phase.

\section{Duality between fragile topological phases and obstructed phases}
\label{sec:duality}
The examples discussed in the previous sections have revealed an intriguing connection between fragile topological phases and obstructed trivial phases: they frequently seem to be related by particle-hole conjugation. In this section, we will show that indeed there is a general connection. First, however, we will need to isolate exactly what kinds of obstructed trivial phases this duality should hold for. For example, the particle-hole conjugate of the Su-Shriefer-Heeger (SSH) chain~\cite{PhysRevLett.42.1698} discussed in Appendix~\ref{sec_mildly_obstructed} is another SSH chain and does not have fragile topology. In this section, we introduce a distinction between \emph{mildly} and \emph{strongly} obstructed trivial phases, and show that the particle-hole conjugate of a \emph{strongly} obstructed trivial phases are always fragile topological. We will phrase the argument for interacting systems, but similar arguments can be applied to non-interacting fermions, so this duality also applies at the level of the non-interacting classification.

For the purpose of this section, we define a deformation between states to be acting with symmetric local unitaries and adding uncharged ancillas (we do not require electronic ancillas). We write $|\Psi\rangle \sim |\Phi\rangle$ to show that the states $|\Psi\rangle$ and $|\Phi\rangle$ can be deformed into each other. A state $|\Psi\rangle$ is fragile* topological if there exists no deformation to an \emph{electronic} atomic insulator (that is, a n atomic insulator where the Hilbert space of each site contains only electronic states, as defined in Section \ref{sec:hilbert}). We will use a subscript $+$ to refer to electronic atomic insulators, e.g. ``$|\phi_+\rangle$''. Note that fragile* topological implies fragile topological in the sense defined earlier, but the converse is not clear; for example, as discussed in Sec.~\ref{sec_breakdown}, three copies of the inversion-protected FTI on the square lattice is not fragile* topological, but we do not know if it is fragile topological.

Now we introduce our notions of obstructed. First of all, we note that the notion of ``obstructed'' depends on specifying a set of ``allowed'' orbitals at ``allowed'' locations in space. Let $\ket{\mathcal{I}}$ be the state that completely fills all the allowed orbitals (which is a particular case of an electronic atomic insulator). A state $|\Psi\rangle$ is (mildly) obstructed trivial if it can be deformed to an electronic atomic insulator, but cannot be deformed to an electronic atomic insulator $|\phi_+\rangle$ which doesn't occupy orbitals that are not in the set of allowed orbitals.
 However, we say $|\Psi\rangle$ is \emph{strongly} obstructed trivial if it satisfies a stronger condition: $|\Psi\rangle$ is strongly obstructed if it is mildly obstructed with respect to \emph{any} set of orbitals which is \emph{deformable} to the original set, in the sense that the fully filled state $\ket{\mathcal{I}'}$ is deformable into $\ket{\mathcal{I}}$. An example of a mildly obstructed phase that is not strongly obstructed is the SSH chain, which is no longer mildly obstructed when we symmetrically move the positions of the ions (which corresponds to a deformation of $\ket{\mathcal{I}}$), as shown in Appendix \ref{sec_mildly_obstructed}.

Let us give an alternative characterization of strongly obstructed trivial: a state $|\Psi\rangle$, which is deformable to an electronic atomic insulator, is strongly obstructed trivial if there is no electronic atomic insulator $|\psi_+\rangle$ such that $|\Psi\rangle \otimes |\psi_+\rangle \sim \ket{\mathcal{I}}$.  Indeed, if there were such an atomic insulator, then given that we know that $|\Psi\rangle \sim |\phi_+\rangle$ for some electronic atomic insulator $|\phi_+\rangle$, we can define $\ket{\mathcal{I}'} := |\phi_+\rangle \otimes |\psi_+\rangle$, and clearly $|\Psi\rangle$ is not mildly obstructed with respect to $\ket{\mathcal{I}'} \sim \ket{\mathcal{I}}$. On the other hand, if $|\Psi\rangle$ is not strongly obstructed, then there exists $\ket{\mathcal{I'}} \sim \ket{\mathcal{I}}$ such that $|\Psi\rangle$ is not mildly obstructed with respect to $\ket{\mathcal{I}'}$, which means that $|\Psi\rangle$ is deformable to an electronic atomic insulator $|\phi_+\rangle$ such that $|\phi_+\rangle$ fills only orbitals that are filled in $\ket{\mathcal{I}'}$. Then it is easy to see that there must be an atomic insulator $|\psi_+\rangle$ such that $|\phi_+\rangle \otimes |\psi_+\rangle \sim \mathcal{I}' \sim \mathcal{I}$.

Now, the particle-hole conjugate is also defined with respect to the ``allowed orbitals''; in particular, at least in the case of states that are not stably topological, one can show that $|\Psi\rangle \otimes |\overline{\Psi}\rangle \sim \ket{\mathcal{I}}$, where $|\overline{\Psi}\rangle$ is the particle-hole conjugate of $|\Psi\rangle$.

Now we can show that if $|\Psi\rangle$ is strongly obstructed, then $|\overline{\Psi}\rangle$ is fragile* topological. To see this, suppose that there were an electronic atomic insulator $|\phi_+\rangle$ such that $|\overline{\Psi}\rangle \sim |\phi_+\rangle$. Then we find that $|\phi_+\rangle \otimes |\Psi\rangle \sim \ket{\mathcal{I}}$. But this exactly contradicts our assumption that $|\Psi\rangle$ was strongly obstructed.

We can also answer the question, if $|\Psi\rangle$ is fragile* topological (but not stably topological), then what is the nature of its particle-hole conjugate $|\overline{\Psi}\rangle$? It turns out that $|\overline{\Psi}\rangle$ must either be strongly obstructed trivial \emph{or} fragile* topological. Indeed, observe that if $|\Psi\rangle$ is not stably topological, then clearly neither is $|\overline{\Psi}\rangle$, so we write $|\overline{\Psi}\rangle \sim |\psi\rangle$ for some atomic insulator $|\psi\rangle$ (possibly involving positrons).
Suppose that $|\overline{\Psi}\rangle$ is not fragile* topological nor strongly obstructed trivial, then there exists an electronic atomic insulator $|\phi_+\rangle$ such that $|\overline{\Psi}\rangle \otimes |\phi_+\rangle \sim \ket{\mathcal{I}}$. We then find that $|\Psi\rangle \otimes \ket{\psi} \sim |\Psi\rangle \otimes |\overline{\Psi}\rangle \sim \ket{\mathcal{I}} \sim |\overline{\Psi}\rangle \otimes |\phi_+\rangle \sim |\psi\rangle \otimes |\phi_+\rangle$. Now we use the fact that for any atomic insulator $\ket{\phi}$ there is an inverse atomic insulator $|\phi^{-1}\rangle$, such that $|\phi\rangle \otimes |\phi^{-1}\rangle = 0$, to conclude that $|\Psi\rangle \sim |\phi_+\rangle$, which contradicts our assumption that $|\Psi\rangle$ was fragile* topological.

\section{Conclusions}
\label{sec:conclusion}
In this work, we prove that certain fermionic fragile topological phases are stable against the introduction of interactions. The gist of the arguments relies on the introduction of fictitious ``positrons''---particles with the opposite charge compared with the physical ones---and by showing that an entanglement-free ground state is possible if and only if positrons are admitted. Our argument extends to problems of interacting bosons, and we construct an example of fragile topological phases for hardcore bosons living on the honeycomb lattice. 

Our results clarify that fragile topological phases can exist in a much more general setting than systems of noninteracting electrons. Their necessary ingredients appear to be spatial symmetries which are rich enough to protect distinct product states~\cite{Zak2000, Oshikawa_SPt, Combinatorics, NC, Bradlyn17}, together with particles carrying a charge which is unbounded and single-signed. Provided oppositely charged particles are physically prohibited, fragile topological phases showcase protected ground-state entanglement much like their conventional counterparts. It remains an interesting open question whether or not such entanglement can lead to nontrivial physical properties. Conversely, when only quantum entanglement is available as a diagnostic on the nontriviality of a state, our discussion implies one should carefully study the stability of such entanglement signatures in the presence of ancillas in order to fully understand the topological nature of the state.\\

\noindent{\it Note added:} In finalizing this manuscript, we received a related manuscript (Ref.~\cite{Ashvin}), which discussed how unconventional physical responses can arise in a system with only fragile, instead of stable, topology. They also discussed the classification of atomic insulators in the presence of interactions.  In places where our discussions overlap, the results are apparently consistent.

\begin{acknowledgements}
HW and DVE would like to acknowledge helpful discussions with Ehud Altman, Xie Chen, Meng Cheng, Michael Hermele, and Ying Ran. DVE also thanks Brayden Ware for helpful discussions.
We also thank Shang Liu, Ashvin Vishwanath, and Eslam Khalaf for sharing their manuscript with us. This work was initiated and performed in part at Aspen Center for Physics, which is supported by National Science Foundation grant PHY-1607611. This research was also supported in part by the National Science Foundation under Grant No.~NSF PHY-1748958.
DVE acknowledges support from the Gordon and Betty Moore Foundation.
HCP is supported by a Pappalardo Fellowship at MIT.
HW acknowledges support from JSPS KAKENHI Grant Number JP17K17678. 
\end{acknowledgements}

\appendix

\section{Equivalence relations beyond homotopy}
\label{appendix_equivalencebeyondhomotopy}
\begin{figure}
\includegraphics[width=0.99\columnwidth]{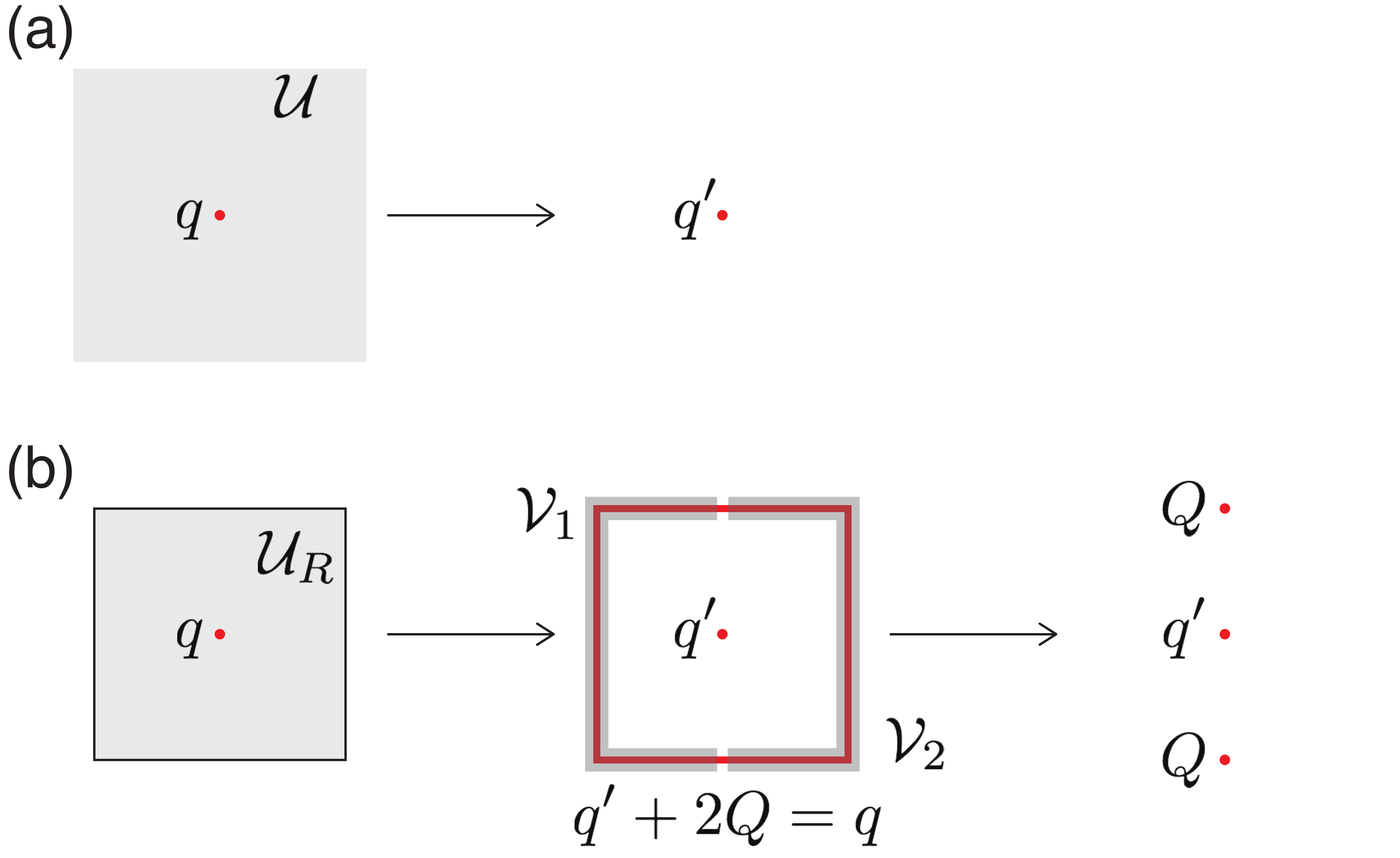}
\caption{(a) If two atomic insulators with charges $q$ and $q'$ at the origin respectively can be related to by a local unitary $\mathcal{U}$ on the plane, then (b) the restriction of $\mathcal{U}$ to a compact region $R$ produces a charge $q'$ at the origin as well as an entangled state on the boundary. This can be symmetrically disentangled in turn.}
\end{figure}

In this section, we will show that the homotopy classification of atomic insulators in two spatial dimensions with respect to inversion symmetry about one point (and no translation symmetry) is correct, in the sense that it generates the same equivalence relatoin as deformations by local unitaries. In two dimensions, inversion is the same as a $C_2$ rotation symmetry; similar arguments will apply to the case of a $C_N$ rotation symmetry, and in fact to any point group in two dimensions, i.e.\ a group of spatial symmetries that leaves one particular point in space invariant. As an example of what we want to show, with a $C_2$ symmetry the $\mathrm{U}(1)$ charge pinned to the rotation center is invariant modulo 2 in atomic insulator homotopies; we want to show that in fact, this charge cannot be changed modulo 2 by \emph{any} local unitary respecting $\mathrm{U}(1)$ and $C_2$ symmetry.

Indeed, consider two atomic insulator states $\ket{\psi}$ and $\ket{\phi}$ with charges $q$ and $q'$ at the origin respectively.
Suppose there exists a symmetric local unitary $\mathcal{U}$ that relates $\ket{\psi}$ to $\ket{\phi}$. By definition, this means there exists a path of local symmetry-respecting Hamiltonians $\mathcal{H}(s)$, $0 \leq s \leq 1$, such that
\begin{equation}
\mathcal{U} = \mathcal{T} \exp\left(-i\int_0^1 \mathcal{H}(s) \right).
\end{equation}

Now consider a large compact $C_2$-symmetric region $R$.
 We can define~\cite{Else_1409} a symmetric restriction $\mathcal{U}_R$ of $\mathcal{U}$ to the region $R$, which acts like $\mathcal{U}$ in the interior of $R$ and like the identity outside of $R$. To define $\mathcal{U}_R$, we just keep all the terms in the Hamiltonian $\mathcal{H}(s)$ which don't act outside of $R$ and throw out the rest. On the region $R$, we can talk about the \emph{global} $\mathrm{U}(1)$ charge of any state. For atomic insulators in particular, we can also talk about the charge at the origin. Moreover, for atomic insulators, since they are product states we can define restrictions $\ket{\psi_R}$ and $\ket{\phi_R}$.

Now consider the effect of acting with $\mathcal{U}_R$ on the state $\ket{\psi_R}$. Obviously, $\mathcal{U}_R \ket{\psi_R}$ must look like $\ket{\phi}$, and hence $\ket{\phi_R}$, on the interior of $R$. However, $\mathcal{U}_R \ket{\psi_R}$ could still carry an entangled one-dimensional state along the boundary of $R$.
But now, following Ref.~\onlinecite{Song_1604}, we consider two non-overlapping regions $Q_{1,2}$ along the boundary of $R$, separated by a distance large compared to the correlation length and related by $C_2$ symmetry. Within $Q_1$, there must be a $\mathrm{U}(1)$-symmetric local unitary $\mathcal{V}_1$ which turns $\mathcal{U}_R \ket{\psi_R}$ to a product state (atomic insulator) within the region [because there are no non-trivial bosonic or fermionic $\mathrm{U}(1)$ SPT phases in (1+1)-D]. By applying the $C_2$-related local unitary to $\mathcal{U}_R \ket{\psi_R}$ in $Q_2$, we now obtain a $C_2$ symmetric atomic insulator. This state has global charge $q$, but the charge at the origin is $q'$. Any charges away from the origin must come in $C_2$-related pairs, so it follows that $q = q' [\operatorname{mod} 2]$.
Finally, let us note that the above arguments are unchanged apply if we allow ancillas, provided that they are brought in in states carrying no charge.

The generalization to other point groups, and to, for example, the inversion eigenvalue at the origin, should be clear. The main point is that if atomic insulators $\ket{\psi}$ and $\ket{\phi}$ are symmetrically deformable into each other, then we can symmetrically deform $\ket{\phi_R}$ into an atomic insulator on $R$ that carries the same charge and inversion eigenvalue at the origin as $\ket{\phi}$, but the same \emph{global} charge and inversion eigenvalue as $\ket{\psi}$. The difference must come from the charges away from the origin, and their contribution is precisely captured by the considerations in Appendix \ref{appendix_fermionvsboson}.

\section{Fermions v.s. Bosons}
\label{appendix_fermionvsboson}
\subsection{Fermions}
Here we provide more details on the deformation process discussed in Sec.~\ref{sec_lattice_homotopy}.  Let us consider an inversion symmetric two-fermion state:
\begin{equation}
|\vec{x},-\vec{x}\rangle_F=\hat{f}_{+\vec{x}}^\dagger\hat{f}_{-\vec{x}}^\dagger|0\rangle,
\end{equation}
where we assume $\hat{f}_{-\vec{x}}^\dagger=\hat{I}\hat{f}_{+\vec{x}}^\dagger\hat{I}^{\dagger}$ is the inversion image of $\hat{f}_{+\vec{x}}^\dagger$ (this amounts to assuming that the inversion symmetry satisfies $\hat{I}^2 = 1$; if $\hat{I}^2 = (-1)^F$ then operators must pick up some extra signs under $\hat{I}$ and the below discussion must be modified.) Due to the fermionic statistics, the inversion eigenvalue of $|\vec{x},-\vec{x}\rangle$ is $-1$:
\begin{equation}
\hat{I}|\vec{x},-\vec{x}\rangle_F=\hat{f}_{-\vec{x}}^\dagger\hat{f}_{+\vec{x}}^\dagger|0\rangle=-\hat{f}_{+\vec{x}}^\dagger\hat{f}_{-\vec{x}}^\dagger|0\rangle=-|\vec{x},-\vec{x}\rangle_F.
\end{equation}
We can rewrite the state $|\vec{x},-\vec{x}\rangle_F$ as
\begin{eqnarray}
|\vec{x},-\vec{x}\rangle_F&=&\hat{p}^\dagger\hat{s}^\dagger|0\rangle\\
\hat{s}&\equiv&\tfrac{1}{\sqrt{2}}(\hat{f}_{+\vec{x}}+\hat{f}_{-\vec{x}}),\\
\hat{p}&\equiv&\tfrac{1}{\sqrt{2}}(\hat{f}_{+\vec{x}}-\hat{f}_{-\vec{x}}).
\end{eqnarray}
This is precisely what Fig.~\ref{kagome4}(h) means. Namely, a pair of fermions at $\pm\vec{x}$ can be understood as the combination of one electron in the $s$-orbital and the other in the $p$-orbital. 

\subsection{Bosons}
Similarly, for bosons,
\begin{equation}
|\vec{x},-\vec{x}\rangle_B=\hat{b}_{+\vec{x}}^\dagger\hat{b}_{-\vec{x}}^\dagger|0\rangle,
\end{equation}
where $\hat{b}_{-\vec{x}}^\dagger=\hat{I}\hat{b}_{+\vec{x}}^\dagger\hat{I}^{\dagger}$. This time, because of the bosonic statistics, we have
\begin{equation}
\hat{I}|\vec{x},-\vec{x}\rangle_B=\hat{b}_{-\vec{x}}^\dagger\hat{b}_{+\vec{x}}^\dagger|0\rangle=\hat{b}_{+\vec{x}}^\dagger\hat{b}_{-\vec{x}}^\dagger|0\rangle=|\vec{x},-\vec{x}\rangle_B.
\end{equation}
In terms of $s$ and $p$ orbitals, we have
\begin{eqnarray}
|\vec{x},-\vec{x}\rangle_B&=&\tfrac{1}{2}[(\hat{s}^\dagger)^2-(\hat{p}^\dagger)^2]|0\rangle,\\
\hat{s}&\equiv&\tfrac{1}{\sqrt{2}}(\hat{b}_{+\vec{x}}+\hat{b}_{-\vec{x}}),\\
\hat{p}&\equiv&\tfrac{1}{\sqrt{2}}(\hat{b}_{+\vec{x}}-\hat{b}_{-\vec{x}}).
\end{eqnarray}
Therefore, unlike Fig.~\ref{kagome4}(h) for fermions, an inversion pair of bosons reduces to the superposition of two $s$-orbitals and two $p$-orbitals.

This difference in the deformation rule results in the difference of the classification of product states in two dimensions symmetric under the inversion and the translation.
For bosons, we get $\mathbb{Z}\times(\mathbb{Z}_2)^3\times(\mathbb{Z}_2)^4$, where $\mathbb{Z}$ is the total filling, $(\mathbb{Z}_2)^3$ is the U(1) charge mod 2 at sites $i=2,3,4$, and $(\mathbb{Z}_2)^4$ is the parity eigenvalue of the sites $i=1,\ldots,4$. 

\subsection{Multi-dimensional irrep of a site symmetry}
In Sec.~\ref{sec_ancillas}, we constructed $G$-symmetric product states using one-dimensional representations of site symmetry $G_x$. Here, we discuss how one can get a one-dimensional representation starting from a multi-dimensional single-particle representation.

Suppose that $u(g)$ is a $D$-dimensional irrep ($D\geq2$) of a site symmetry. Let $\{f_i^\dagger\}_{i=1}^D$ be a fermionic multiplet satisfying
\begin{equation}
\hat{g}\hat{f}_i^\dagger\hat{g}^\dagger=\hat{f}_j^\dagger U(g)_{ji}.
\end{equation}
Since these $D$ single-particle modes transform into one another under the symmetries, a way to construct a symmetric state is to fill all of them. As such, one finds that
the $D$-particle state $|D\rangle=\hat{f}_1^\dagger\hat{f}_2^\dagger\ldots\hat{f}_D^\dagger|0\rangle$ obeys the one-dimensional representation
\begin{equation}
\hat{g}|D\rangle=\text{det}[U(g)]|D\rangle.
\end{equation}
 Note, however, that this is not the only way to construct a one-dimensional representation for fermions. While the state $|D\rangle$ above is still a Slater determinant, in general one can construct inherently interacting states through the linear superpositions of Slater determinants. Such states could furnish symmetry representations that are not realizable within the free-fermion states~\cite{FragileMott}. 
 
\section{The SSH model}
\label{sec_mildly_obstructed}
As the canonical example of mildly obstructed trivial phases introduced in Sec.~\ref{sec:duality}, let us discuss the SSH chain~\cite{PhysRevLett.42.1698}.  The symmetry group $G$ is generated by the U(1) symmetry, the lattice translation symmetry, and the inversion symmetry. In each unit cell $R\in\mathbb{Z}$, we assume one site at $x=R+\xi$ and the other at $x=R-\xi$ ($0< \xi<0.5$). The model is thus defined on a lattice $\Lambda=\{R\pm\xi\,|\,R\in\mathbb{Z}\}$.  The tight-binding model in the Fourier space reads $H_{k}=(t+t'\cos k)\sigma_1+t'\sin k\sigma_2$, where $t$ and $t'$ are the intra-cell and the inter-cell hopping.  The inversion symmetry is represented by $\sigma_1$.  

When $t\neq t'$, there exists a nonzero band gap. Let $|\Psi\rangle_{\Lambda}$ be the insulating ground state that fully occupies the lower band.
Suppose that the Wannier center of the lower band locates at $x=R\in\mathbb{Z}$ (it may as well be $x=R+\frac{1}{2}$ and the same argument applies to this case). Then, $|\Psi\rangle_{\Lambda}$ would be deformable to a product state in which charges are strictly localized at every $x\in\mathbb{Z}$, but such a state does not belong to our original Hilbert state $\mathcal{H}_{\Lambda}$. However, this obstruction can be resolved by smoothly changing the value of $\xi$ to 0. Importantly, no ancillas are required in performing this deformation.

\bibliography{reference}
\end{document}